\documentclass[journal]{IEEEtran}

\usepackage{array}
\usepackage{amsmath,amssymb,amsfonts}
\usepackage{textcomp}
\usepackage{url}
\usepackage{verbatim}
\usepackage{graphicx}
\usepackage{cite}
\usepackage{xcolor}
\usepackage{algorithm}
\usepackage{algorithmicx}
\usepackage{algpseudocode}
\usepackage{epstopdf}
\usepackage{bm}
\usepackage{gensymb}
\usepackage{svg}
\usepackage{mathrsfs}
\usepackage{caption}
\usepackage{subcaption}
\usepackage{float}
\usepackage{stfloats}

\newcommand{\mv}[1]{\mathbf{#1}}
\newcommand{\mm}[1]{\mathbf{#1}}
\newcommand{\mtr}{\mathrm{tr}}

\newcommand{\mstr}[1]{\mathrm{#1}}

\usepackage{hyperref}
\begin{document}

\title{Passive Detection in Multi-Static ISAC Systems: Performance Analysis and Joint Beamforming Optimization}
\author{Renjie He, Yiqiu Wang, Meixia Tao,~\IEEEmembership{Fellow,~IEEE}, and Shu Sun,~\IEEEmembership{Senior Member,~IEEE}
\thanks{The authors are with the Department of Electronic Engineering and the Cooperative Medianet Innovation Center (CMIC), Shanghai Jiao Tong University, China (e-mails: \{renjiehe\_sjtu, wyq18962080590, mxtao, shusun\}@sjtu.edu.cn).}
\thanks{Part of this work was presented in the 10th workshop on integrated sensing and communications for edge Intelligence at IEEE Global Telecommunications Conference (GLOBECOM), Cape Town, South Africa, December 12, 2024 \cite{previouswork}.}
}

\maketitle

\begin{abstract}
This paper investigates the passive detection problem in multi-static integrated sensing and communication (ISAC) systems, where multiple sensing receivers (SRs) jointly detect a target using \textit{random unknown} communication signals transmitted by a collaborative base station. Unlike traditional active detection, the considered passive detection does not require complete prior knowledge of the transmitted communication signals at each SR. First, we derive a generalized likelihood ratio test detector and conduct an asymptotic analysis of the detection statistic under the large-sample regime. We examine how the signal-to-noise ratios (SNRs) of the target paths and direct paths influence the detection performance. Then, we propose two joint transmit beamforming designs based on the analyses. In the first design, the asymptotic detection probability is maximized while satisfying the signal-to-interference-plus-noise ratio requirement for each communication user under the total transmit power constraint. Given the non-convex nature of the problem, we develop an alternating optimization algorithm based on the quadratic transform and semi-definite relaxation. The second design adopts a heuristic approach that aims to maximize the target energy, subject to a minimum SNR threshold on the direct path, and offers lower computational complexity. Numerical results validate the asymptotic analysis and demonstrate the superiority of the proposed beamforming designs in balancing passive detection performance and communication quality. This work highlights the promise of target detection using unknown communication data signals in multi-static ISAC systems.
\end{abstract}

\begin{IEEEkeywords}
Integrated sensing and communication, passive radar detection, generalized likelihood ratio test, multi-static, beamforming design.
\end{IEEEkeywords}

\section{Introduction}
\IEEEPARstart{T}{he} sixth-generation (6G) wireless network is envisioned to support a broad range of advanced applications, such as autonomous driving, smart cities, and digital twins, which require both reliable communication and high-precision sensing capabilities. Integrated sensing and communication (ISAC) technology has emerged as a promising technology to efficiently meet these demands. By combining sensing and communication functions into a unified system with shared hardware platforms, spectra, and waveforms, ISAC can significantly improve resource utilization and facilitate mutual performance gains \cite{gonzalez2024integrated, liu2022integrated}. This synergy has made ISAC a pillar usage scenario in 6G and necessitates collaborative innovations across radar and communication domains \cite{ITU}.

\begin{figure}
    \centering
    \begin{subfigure}[b]{0.24\textwidth}
        \centering
        \includegraphics[width=\textwidth]{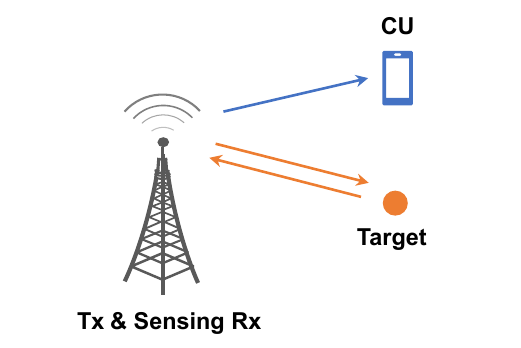}
        \caption{}
        \label{fig:subfig1}
    \end{subfigure}
    \hfill  
    \begin{subfigure}[b]{0.24\textwidth}
        \centering
        \includegraphics[width=\textwidth]{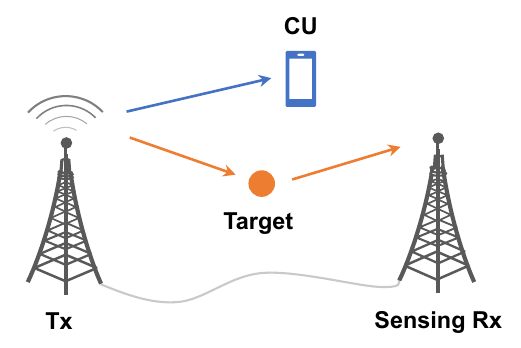}  
        \caption{}
        \label{fig:subfig2}
    \end{subfigure}
    
    \vspace{1em}  
    
    \begin{subfigure}[b]{0.24\textwidth}
        \centering
        \includegraphics[width=\textwidth]{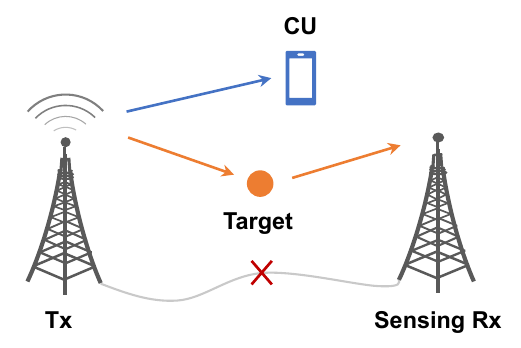}  
        \caption{}
        \label{fig:subfig3}
    \end{subfigure}
    \hfill
    \begin{subfigure}[b]{0.24\textwidth}
        \centering
        \includegraphics[width=\textwidth]{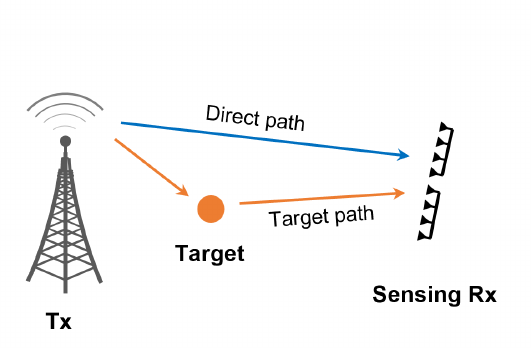}  
        \caption{}  
        \label{fig:subfig4}
    \end{subfigure}
    
    \caption{System illustrations. (a): Mono-static ISAC system; (b): Bi-static ISAC system with known transmitted signal; (c): Bi-static ISAC system with unknown transmitted signal; (d): Bi-static passive radar system with unknown transmitted signal and a two-channel SR.}  
    \label{fig:example}  
    \vspace{-0.3cm}
\end{figure}

ISAC systems implement sensing functionality through two primary configurations, mono-static sensing and bi/multi-static sensing, depending on whether the sensing receiver (SR) is co-located with or spatially separated from the transmitter. Mono-static sensing typically requires a full-duplex transceiver to simultaneously transmit signals and receive target echoes (as shown in Fig. \ref{fig:subfig1}), and therefore may pose substantial challenges to self-interference cancellation \cite{xiao2022waveform}. In contrast, bi/multi-static sensing employs one or more spatially distributed SRs (as shown in Fig. \ref{fig:subfig2},(c)), thus avoiding the self-interference issue and enhancing the sensing coverage, which is of significant practical interest \cite{zhang2021enabling}. However, the physical separation also poses new challenges, particularly in designing collaboration frameworks for dual purposes, which have garnered considerable interest in recent studies.

Current multi-static ISAC systems predominantly rely on collaborative mechanisms between transmitters and SRs. It is typically assumed that SRs possess full prior knowledge of the sensing signals to enable subsequent signal processing, which is also known as active sensing. This can be achieved by transmitting predefined sensing signals or by sharing signal information with SRs through dedicated links, as illustrated in Fig. \ref{fig:subfig2}. Predefined sensing signals or reference signals from existing physical-layer protocols are convenient choices to facilitate sensing, but they compete with communication signals for time-frequency resources, and potentially cause interference with communication users (CUs) \cite{dong2022sensing}. Additionally, reference signals typically occupy only about 10\% of time-frequency resources \cite{lu2024random}, which limits their utility for high-precision and continuous sensing. To improve resource utilization and enhance sensing capabilities in multi-static ISAC systems, reusing communication data signals for sensing purposes has attracted increasing attention. In \cite{xiong2023fundamental}, the fundamental performance trade-off between sensing and communication with \textit{random known} signals is characterized by analyzing the Cram\'er-Rao bound-rate region. In \cite{liu2025sensing, li2023toward, cheng2024optimal}, SRs obtain communication data signals from the ISAC transmitter via wired links to construct matched filters for sensing. In these studies, complete prior knowledge of the communication signal is required at each SR for sensing signal processing. However, this introduces substantial signaling overhead and time delay. Moreover, when SRs are external to the wireless network infrastructure, obtaining such prior information requires additional protocols, thereby increasing the system complexity.

These practical issues have resulted in a notable sensing paradigm where \textit{random unknown} communication signals are used for target sensing (as depicted in Fig. \ref{fig:subfig3}), which is recognized as the passive sensing problem in the radar literature \cite{santamaria2017passive, 9496200, kuschel2019tutorial}. Passive sensing can be implemented using both downlink and uplink communication signals. It is a challenging task to sense targets without complete prior knowledge of both transmit signals and sensing channels, and only limited ISAC literature has investigated this problem. An intuitive strategy is to first demodulate the communication symbols (with the help of pilot symbols) and then use the symbols to reconstruct the transmit signals for sensing. But this approach suffers from performance degradation due to symbol detection errors \cite{9880826}. To overcome this problem, a message passing algorithm is designed for joint data demodulation and environment imaging in an uplink ISAC system in \cite{zhu2023ris}. In \cite{jiang2024isac}, the symbol detection, sensing algorithm and channel reconstruction are iteratively updated in a downlink bi-static ISAC system. The localization performance of such iterative methods is further analyzed in \cite{10718341}. In these studies, the random unknown symbols have to be detected with the assistance of known reference signals, similar to conventional communication systems. Thus, passive sensing is viable only when a communication link is established between the BS and SR. In addition, existing discussions are restricted to bi-static configurations and with a single user data stream, whereas the design and analysis complexity increases significantly as the number of SRs and the number of user data streams increase \cite{steffes2022passive}.

To overcome the limitations of passive sensing in multi-static ISAC with random unknown signals, we can leverage the design insights from the passive radar literature. In passive radar systems, SRs utilize unknown ambient communication signals from non-cooperative transmitters to perform target sensing. A common approach is to employ a two-channel SR \cite{kuschel2019tutorial, 9661315}, as shown in Fig. \ref{fig:subfig4}, where each SR has two arrays, one for surveillance to capture target reflected echoes, which is referred to as target path in this paper, and the other for reference to receive the line-of-sight (LOS) path signals from the transmitter to the receiver, which provides critical reference information about the unknown transmitted signal and is referred as direct path in this paper. The correlations between these two channels are then exploited for sensing \cite{hack2014detection}. Although similar to passive radar, passive sensing in multi-static ISAC has distinct design considerations and challenges. Specifically, passive radar research focuses primarily on advanced receiver design, as the non-cooperative transmitter lies beyond the control of the system \cite{salah2013feasibility}. In contrast, ISAC systems allow for collaborative transmission strategies, such as optimizations on waveform and beamforming for enhanced sensing and communication performance. In addition, passive radar typically utilizes rank-one broadcast signals from omnidirectional transmitters \cite{steffes2022passive}, whereas ISAC systems normally utilize antenna arrays to transmit multi-user signals. The passive detection performance with beamformed target illumination and general rank signals has not been sufficiently discussed in the existing literature. Joint transmit beamforming design holds the potential to improve both passive detection and communication performance, yet has not been thoroughly investigated.

In this paper, we aim to investigate the passive detection problem in a multi-static ISAC system with one multi-antenna base station (BS) and multiple two-channel SRs. Similar to passive radar, one channel from each SR is used to receive target echoes, and the other channel is used to receive signals from the LOS path. We first derive a generalized likelihood ratio test (GLRT) detector for passive detection and analyze its asymptotic performance under the large-sample regime. Then, we formulate two joint beamforming optimization problems based on our analyses. The main contributions and results in this work are summarized as follows:
\begin{itemize}
\item First, we establish a passive detection model in a multi-static multi-input multi-output (MIMO) ISAC scenario, where multiple two-channel SRs perform passive target detection using \textit{random unknown} communication signals. We derive a GLRT detector and analyze its asymptotic distribution based on Wilk's theorem. The analysis demonstrates that asymptotic performance depends on the signal-to-noise ratios (SNRs) of both the direct and target paths, but the SNR of target paths takes a dominant role.

\item Second, we propose two joint transmit beamforming problem formulations and their corresponding algorithms based on our analyses. The first problem aims to maximize the asymptotic detection probability, subject to the transmit power budget and the minimum signal-to-interference-plus-noise ratio (SINR) requirement of all CUs. This problem is highly nonconvex, and thus we utilize the quadratic transform and semidefinite relaxation (SDR) technique to reformulate the objective and propose an alternating optimization algorithm. The second problem is a heuristic one that aims to maximize the target energy while guaranteeing the minimum average SNR in the direct paths. This simplified problem can be solved using SDR without alternating procedure, resulting in lower complexity.

\item Finally, we verify our analysis and beamforming designs through extensive numerical simulations. The asymptotic detection performance is compared with both the empirical and active detection performances. The asymptotic result is shown to be a reasonable approximation to the actual performance when the direct path quality is acceptable. Furthermore, simulations demonstrate that the proposed beamforming designs can effectively improve the detection probability compared to the benchmark schemes.

\end{itemize}

The remainder of this paper is organized as follows. In Section II, we introduce the ISAC system model, including the communication and passive detection model. In Section III, we derive asymptotic analyses of the passive detection performance. In Section IV, we elaborate on the joint ISAC beamforming optimization problem and the corresponding algorithms. Section V provides the numerical results. Finally, Section VI concludes the paper.

\textit{Notations:} $[\cdot]^T$, $[\cdot]^*$, $[\cdot]^H$ denote the transpose, conjugate, and Hermitian transpose of a matrix, respectively; $\mathbb{E}[\cdot]$ denotes the expectation of a random variable; $\Re[\cdot]$, $\Im[\cdot]$ denote the real and imaginary part of a complex number, respectively; $\mm A \succeq 0$ denotes a semi-definite matrix $\mm A$; $|\cdot|$ denotes the magnitude of a complex number; $||\cdot||$ denotes the Euclidean norm; $\mstr{vec}(\cdot)$ is the vectorization operator; $\mm 1(a\geq b)$ is an indicator function which equals to $1$ if $a\geq b$ or $0$ if otherwise; $\mm I_{n}$ denotes an $n\times n$ identity matrix; $\mm 0_{m\times n}$ denotes an $m\times n$ zero matrix. 

\section{System Model} \label{sec:System Model}
\begin{figure}
    \centering
    \includegraphics[width=0.5\textwidth]{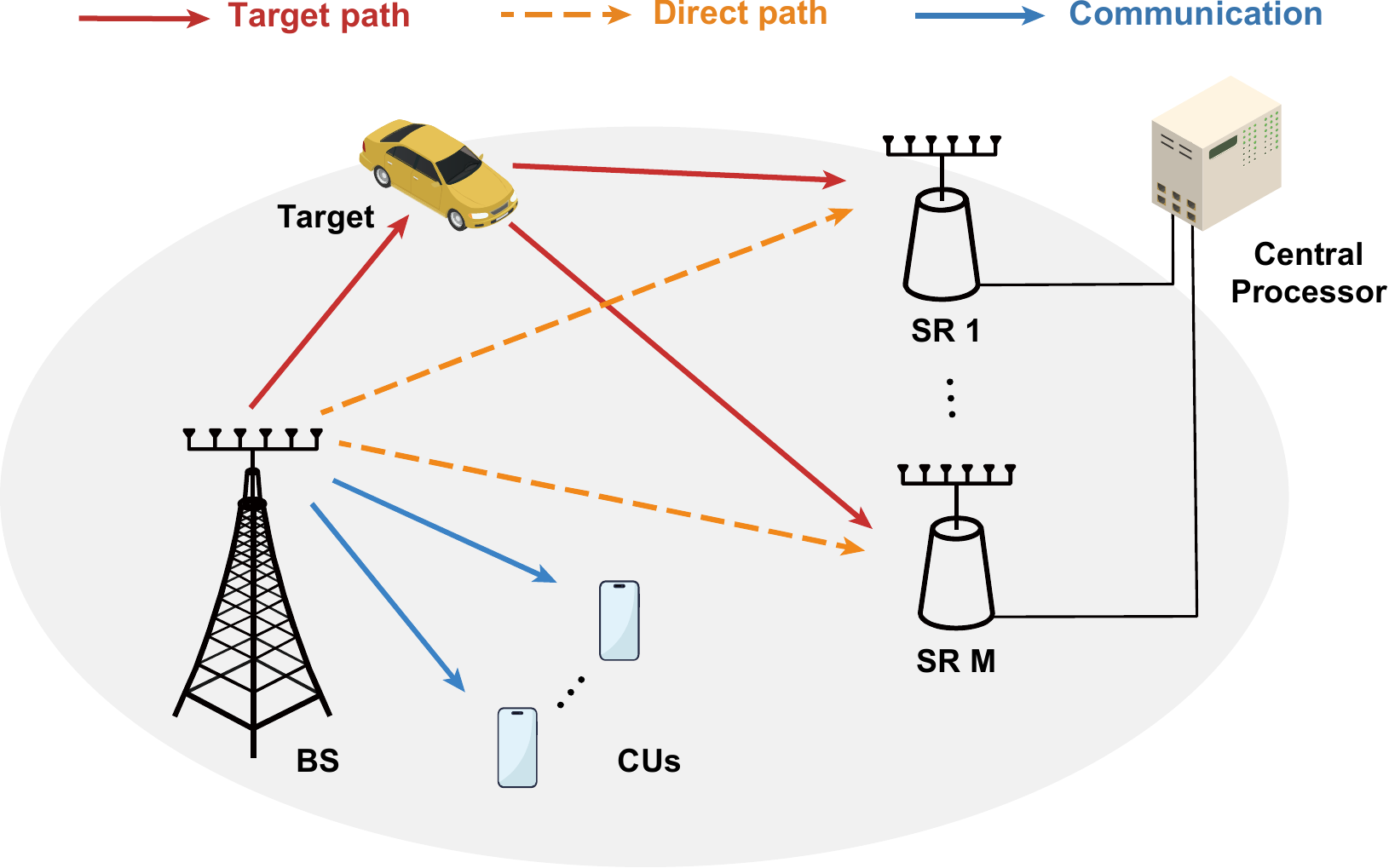}
    \caption{Diagram of the considered ISAC system.}
    \label{fig:system}
\end{figure}

\begin{figure}
    \centering
    \includegraphics[width=0.5\textwidth]{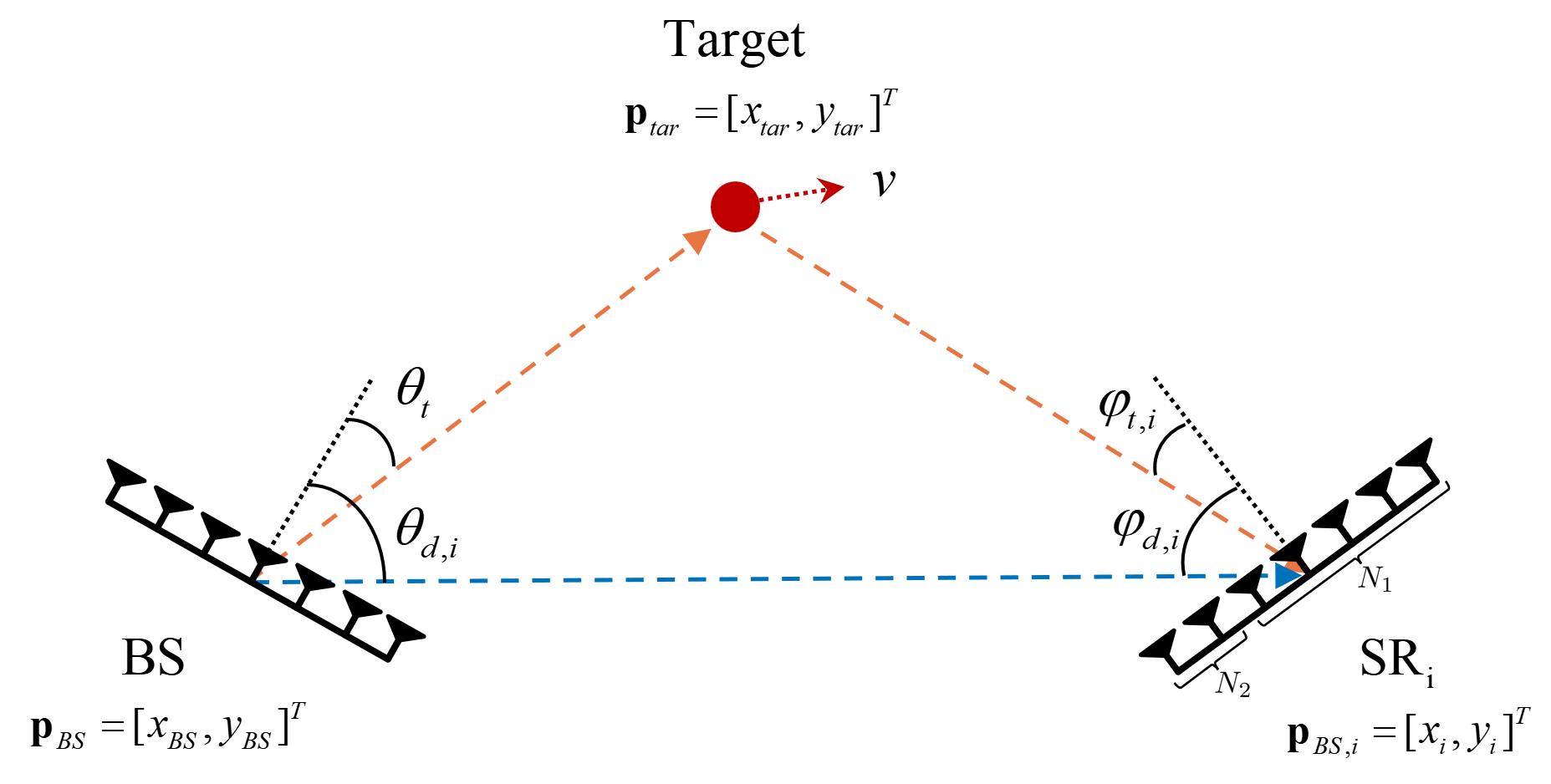}
    \caption{Geometry illustration.}
    \label{fig:geometryillus}
\end{figure}

As shown in Fig. \ref{fig:system}, we consider a multi-static ISAC scenario, where a BS serves $C$ downlink CUs and $M$ SRs simultaneously. To employ random unknown communication signals for passive detection, each SR is equipped with a classic two-channel passive radar receiver, consisting of a surveillance array with $N_1$ antennas and a reference array with $N_2$ antennas, as shown in Fig. \ref{fig:geometryillus}. The surveillance array focuses on extracting target echoes along the so-called target path, while the reference array collects direct path signals. Each SR sends locally processed signals to a central processor via dedicated links for joint detection. Suppose that each CU has a single antenna and the BS is equipped with a uniform linear array with $N_t$ antenna elements, where $N_t \geq C$. In the following, we first establish the downlink communication model, then introduce the multi-static passive detection model and derive a GLRT detector.

\subsection{Communication Model}

Let $s_n[l]$ denote the $l$-th communication data symbol intended to the $n$-th CU, for $l \in \{1, \cdots, L\}$ with $L$ being the data block length. The transmitted baseband signal can be written as:
\begin{equation}
    \mm x[l] = \sum_{n = 1}^C \mm w_n s_n[l],
\end{equation}
where $\mm w_n \in \mathbb{C}^{N_{t}}$ is the transmit beamforming vector for the $n$-th CU, and $\{\mm w_n\}_{n=1}^C$ are jointly designed by considering both communication and sensing performance. Let $\mm s[l] = [s_1[l], \cdots, s_C[l]]^T$ denote the symbol vector, which is independent and identically distributed (i.i.d.) and follows the circularly symmetric complex Gaussian distribution with zero mean and unit power, that is, $\mm s[l] \sim \mathcal{CN}(0, \mm I_C)$. Then, the covariance matrix of the transmitted signal is given by:
\begin{equation}
    \mm R_{\text{c}} = \mathbb{E}[\mm x[l] \mm x^H[l]] = \mm W\mm W^H,
\end{equation}
where $\mm W = [\mm w_1, \cdots, \mm w_C]$ is the beamforming matrix. We assume the communication channel $\mm h_m \in \mathbb{C}^{ N_{t}}$ remains constant within the considered time interval and is perfectly known by the BS. The received signal $r_m[l]$ at the $m$-th CU is given by:
\begin{equation}
    r_m[l] = {\mm h_m^H} \sum_{n = 1}^C \mm w_n s_n[l] + n_m[l],
\end{equation}
where $n_m[l] \sim \mathcal{CN}(0, \sigma_{\text{c}}^2)$ is the additive white Gaussian noise (AWGN) with zero-mean and variance of $\sigma_{\text{c}}^2$. To guarantee communication quality, the SINR of each CU should exceed a certain threshold:
\begin{equation}
    \gamma_n = \frac{|\mm h_n^H\mm w_n|^2}{\sigma_{\text{c}}^2 + \sum_{k \neq n, k = 1}^C |\mm h_k^H\mm w_k|^2} \geq \Gamma_{\text{c}, n}, \quad \forall n.
\end{equation}

\subsection{Passive Detection Model} \label{sec: II-B}
In this sub-section, we first establish the sensing reception model, then present the hypothesis testing problem. 

The BS and the $i$-th SR are located at known positions, denoted as $\mm p_{\text{BS}} = [x_{\text{BS}}, y_{\text{BS}}]^T$ and $\mm p_{\text{SR}, i} = [x_{i}, y_{i}]^T$, respectively. Suppose that there is one point target presenting at $\mm p_{\text{tar}} = [x_{\text{tar}}, y_{\text{tar}}]^T$. As shown in Fig. \ref{fig:geometryillus}, we denote $\theta_t$ as the angle of departure (AOD) of the target, $\theta_{d, i}$ as the AOD of the $i$-th SR, $\varphi_{t, i}$ as the angle of arrival (AOA) of the target at the $i$-th SR, and $\varphi_{d, i}$ as the AOA of direct path signal at the $i$-th SR. Let ${\mm{q}}_N(\theta)$ denote the steering vector of an $N$-element antenna array at angle $\theta$:
\begin{equation}
    \mm{q}_N(\theta) = \left[ 1, e^{j\frac{2\pi}{\lambda}\Delta \sin\theta}, \cdots, e^{j\frac{2\pi}{\lambda}\Delta (N - 1)\sin\theta} \right],
\end{equation}
where $\lambda$ is the wavelength, $\Delta$ is the antenna spacing. For brevity, we denote the transmit steering vectors as $\mm a_t = \mm{q}_{N_t}(\theta_t)$, $\mm a_{d, i} = \mm{q}_{N_t}(\theta_{d, i})$, denote the receive steering vectors of the surveillance array as $\mm b_{t, 1, i} = \mm{q}_{N_1}(\varphi_{t, i})$, $\mm b_{d, 1, i} = \mm{q}_{N_1}(\varphi_{d, i})$, and denote the receive steering vectors of the reference array as $\mm b_{t, 2, i} = \mm{q}_{N_2}(\varphi_{t, i})$, $\mm b_{d, 2, i} = \mm{q}_{N_2}(\varphi_{d, i})$. 

To characterize the time delays and Doppler shifts introduced by the propagation channel over the $L$ symbol durations, we define the delay-Doppler operator $\mm D(\cdot) \in \mathbb{C}^{L \times L}$ as \cite{hack2014detection}:
\begin{equation}
\label{eq:DD}
    \mm D(\tau, f) \triangleq \frac{1}{L}\mm K(f/f_s) \Pi_L^H \mm K(-\tau f_s/L) \Pi_L ,
\end{equation}
where $\tau$ is time delay and $f$ is the Doppler shift, $\Pi_L \in \mathbb{C}^{L \times L}$ is the discrete Fourier transform matrix, $\mm K(x) = \mstr{diag}([1, e^{j 2\pi x}, \cdots, e^{j 2\pi (L-1) x}])$, and $f_s$ is the sampling frequency. Then, for the target and direct paths of the $i$-th SR, the delay-Doppler operators are given by $\mm D_{t, i} = \mm D(\tau_{t, i}, f_{i})$ and $\mm D_{d, i} = \mm D(\tau_{d, i}, 0)$, respectively. Here, $\tau_{t, i} = (||\mm p_{\text{BS}} - \mm p_{\text{tar}}||+||\mm p_{\text{tar}} - \mm p_{\text{SR}, i}||)/c$ and $\tau_{d, i} = ||\mm p_{\text{BS}} - \mm p_{\text{SR}, i}||/c$ are the propagation delays of the target path and direct path at SR $i$, respectively, $c$ is the speed of light, and $f_{i}$ is the Doppler shift introduced by the moving target with respect to SR $i$.

Assume that the target detection is performed over a coherent processing interval with $L$ snapshots, with each snapshot corresponding to one symbol vector. The sampled complex baseband signals received at the $i$-th SR by both the surveillance and the reference arrays can be expressed as: 
\begin{equation}
\label{eq:received}
    \begin{aligned}
        \mm Y_{k, i} = \alpha_{t, i} {\mm b}_{t, k, i} {\mm a}_{t}^H \mm W \mm S \mm D_{t, i}^T + \alpha_{d, i} {\mm b}_{d, k, i} {\mm a}_{d, i}^H \mm W \mm S \mm D_{d, i}^T + \mm Z_{k, i},
    \end{aligned}
\end{equation}
where $\mm Y_{k, i}\in \mathbb{C}^{N_k \times L}$ with $k=1$ and $2$ representing the surveillance and reference array, respectively, $\mm S = [\mm s[1], \cdots, \mm s[L]]$ is the transmitted symbol matrix, $\mm Z_{k, i}\in \mathbb{C}^{ N_{k}\times L}$ is the AWGN matrix, $\alpha_{d, i}$ is the direct path gain of SR $i$, $\alpha_{t, i}$ is the target path gain of SR $i$ that includes the target radar cross section (RCS).

Note that in practical situations, both arrays at each SR may receive not only the target echoes and the direct-path signals, but also clutter. Since there have been extensive techniques for mitigating clutter \cite{bolvardi2017dynamic}, we have focused on a clutter-free signal model in (\ref{eq:received}). The corresponding sensing performance could serve as an upper bound for a more realistic scenario. 

Based on the reception model, we next present the sensing procedure. 

\textit{Step 1: Receive Beamforming.} In general, the direct path signal is much stronger than the target echoes \cite{kuschel2019tutorial}, and thus may mask weak targets and severely degrade the detection performance. Therefore, suppressing the direct-path interference in the surveillance array is necessary for passive sensing, normally by employing temporal or spatial filtering \cite{hack2014detection, zhang2017multistatic}. In this work, we adopt a receive beamforming approach, where the echoes and direct signals are isolated in the surveillance and reference arrays. The corresponding receive beamforming vectors for the $i$-th SR are denoted as $\mm q_{t, i}\in \mathbb{C}^{ N_1}$ and $\mm q_{d, i}\in \mathbb{C}^{ N_2}$, which can be easily calculated through orthogonalization process. The separated target and direct path signals at the $i$-th SR can be expressed as:
\begin{equation}
\label{eq:separation1}
    \begin{aligned}
        \mm Y_{1, i}' &= \mu_{t, i} {\mm a}_{t}^H \mm W \mm S \mm D_{t, i}^T + \mm Z_{1, i}', \\
        \mm Y_{2, i}' &= \mu_{d, i} {\mm a}_{d, i}^H \mm W \mm S \mm D_{d, i}^T + \mm Z_{2, i}', 
    \end{aligned}
\end{equation}
where $\mu_{t, i} = \alpha_{t, i}\mm q_{t, i}^H {\mm b}_{t, 1, i}$ and $\mu_{d, i} = \alpha_{d, i} \mm q_{d, i}^H {\mm b}_{d, 2, i}$ denote the equivalent channel gain after the receive beamforming, $\mm Z_{1, i}'$ and $\mm Z_{2, i}'$ are independent AWGN matrices. 

\textit{Step 2: Delay-Doppler Compensation.} Following a common practice in radar detection, we consider the target detection within a hypothesized position-velocity cell. We discretize the continuous delay and Doppler domain at each SR into a set of tuples $\{(\hat\tau_{i, j}, \hat f_{i, j})\}_{j=1}^{N_0}$, for $i\in \{ 1, \cdots, M\}$. For tuple $(\hat\tau_{i, j}, \hat f_{i, j})$, we apply operator $\left[\mm D(\hat\tau_{i, j}, \hat f_{i, j})\right]^*$ for the delay-Doppler compensation. To simplify the analysis, we assume that there exists a delay-Doppler tuple $j_{t, i}^\star \in \{1, ..., N_0\}$ for the target channel that matches the true target delay and Doppler, that is, $\mm D(\hat\tau_{i, j_{t, i}^\star}, \hat f_{i, j_{t, i}^\star}) = \mm D_{t, i}$ (the impacts of mismatches have been studied in \cite{5417172} and are not our focus here). Similarly, we assume that the delay and Doppler of the direct channel can also be matched, that is, $\mm D(\hat\tau_{i, j_{d, i}^\star}, \hat f_{i, j_{d, i}^\star}) = \mm D_{d, i}$ with $j_{d, i}^\star \in \{1, ..., N_0\}$. Then, the compensated signals can be written as:
\begin{equation}
\label{eq:separation}
    \begin{aligned}
        \tilde{\mm Y}_{1, i} &= \mm Y_{1, i}' \mm D_{t, i}^* = \mu_{t, i} {\mm a}_{t}^H \mm W \mm S + \tilde{\mm Z}_{1, i}, \\
        \tilde{\mm Y}_{2, i} &= \mm Y_{2, i}' \mm D_{d, i}^* = \mu_{d, i} {\mm a}_{d, i}^H \mm W \mm S + \tilde{\mm Z}_{2, i},
    \end{aligned}
\end{equation}
where $\tilde{\mm Z}_{1, i}$ and $\tilde{\mm Z}_{2, i}$ are independent AWGN matrices. 

\textit{Step 3: Signal Aggregation and Detection.} The processed samples from each SR are aggregated at the central processor, which are denoted as $\tilde{\mm Y}_t = [\tilde{\mm Y}_{1, 1}^T, \cdots, \tilde{\mm Y}_{1, M}^T]^T$ and $\tilde{\mm Y}_d = [\tilde{\mm Y}_{2, 1}^T, \cdots, \tilde{\mm Y}_{2, M}^T]^T$. The noise matrices can be similarly defined. Finally, the aggregated signals are fed to a detector. 

Next, we formally introduce the hypothesis test formulation for the considered passive detection problem. For notation simplicity, we define the following equivalent channel matrices:
\begin{equation}
\label{eq:H}
    \tilde{\mm H}_d \triangleq \begin{bmatrix} 
    \mu_{d, 1} {\mm a}_{d, 1}^H \mm W \\ \vdots
    \\ \mu_{d, M} {\mm a}_{d, M}^H \mm W \end{bmatrix}, \quad \tilde{\mm H}_t \triangleq \begin{bmatrix} 
    \mu_{t, 1} {\mm a}_{t}^H \mm W \\ \vdots
    \\ \mu_{t, M} {\mm a}_{t}^H \mm W \end{bmatrix}\in \mathbb{C}^{M\times C}.
\end{equation}

Because these equivalent channel matrices are unknown but deterministic to the SRs, the passive detection problem can be formulated as a two-channel composite hypothesis testing problem, as illustrated below:
\begin{equation}
\label{eq:HT}
    \begin{aligned} 
        \mathcal{H}_0: \quad
        \tilde{\mm Y}_t & = \tilde{\mm Z}_{t},\\
        \tilde{\mm Y}_d & = \tilde{\mm H}_d \mm S + \tilde{\mm Z}_{d}, \\
        \mathcal{H}_1: \quad
        \tilde{\mm Y}_t & = \tilde{\mm H}_t \mm S + \tilde{\mm Z}_{t},\\
        \tilde{\mm Y}_d & = \tilde{\mm H}_d \mm S + \tilde{\mm Z}_{d},\\
    \end{aligned}
\end{equation}
where $\mathcal{H}_0$ is the null hypothesis indicating that no target presents, and $\mathcal{H}_1$ is the alternative hypothesis. Denote the overall concatenated signal as $\mm Y = [\tilde{\mm Y}_t^T, \tilde{\mm Y}_d^T]^T \in \mathbb{C}^{2M \times L}$. In each snapshot, we have $\mm Y[l] \sim \mathcal{CN}(\mm 0, \mm{R}_{y, j})$, where the covariance matrix can be written as:
\begin{align}
\label{eq:covriancemat}
    \mm{R}_{y, j} = \begin{bmatrix} 
    \tilde{\mv H}_t \tilde{\mv H}_t^H + \sigma_r^2 \mm I_M & \tilde{\mv H}_t \tilde{\mv H}_d^H\\ \tilde{\mv H}_d \tilde{\mv H}_t^H & \tilde{\mv H}_d \tilde{\mv H}_d^H + \sigma_r^2 \mm I_M \end{bmatrix},
\end{align}
where $j=0$ refers to $\mathcal{H}_0$ with $\tilde{\mv H}_t = \mm 0$, and $j=1$ refers to $\mathcal{H}_1$ with $\tilde{\mv H}_t \neq \mm 0$. In (\ref{eq:covriancemat}), we have utilized the fact that noises in the target and direct paths are uncorrelated due to the use of two arrays. Besides, $\mm Y[l]$ and $\mm Y[k]$ are independent variables for $l\neq k$. Then, the probability density function (PDF) of the received signal matrix $\mm Y$ can be given by:
\begin{equation}
\label{eq:PDF}
p(\mm Y | \mathcal{H}_j) 
        = \frac{1}{\pi^{2ML} (\det \mm{R}_{y, j})^L} e^{-\mathrm{tr} (\mm Y^H \mm{R}_{y, j}^{-1} \mm Y)},
\end{equation}
where the covariance matrix $\mm R_{y,j}$ is deterministic but unknown. 

\subsection{GLRT Detector}
The GLRT is an effective approach to deal with the composite hypothesis testing problem, which replaces unknown parameters in both the null and alternative distributions with their maximum likelihood estimations (MLEs). Here we need to derive the MLEs of $\tilde{\mv H}_t$ and $\tilde{\mv H}_d$ from (\ref{eq:HT}). The log-statistic of the GLRT can be calculated by:
\begin{equation}
\label{eq:stadef}
\Lambda(\mm Y) = \ln  \frac{\max_{\tilde{\mv H}_t, \tilde{\mv H}_d} p(\mm Y | \mathcal{H}_1)}{\max_{\tilde{\mv H}_d} p(\mm Y | \mathcal{H}_0)} \overset{\mathcal{H}_1} {\underset{\mathcal{H}_0} {\gtrless}} \rho,
\end{equation}
where $\rho$ is the decision threshold corresponding to a predefined false alarm rate. 

\textit{Proposition 1:} The detection statistic $\Lambda(\mm Y)$ is given by:
\begin{align}
\label{eq:statistic}
\Lambda(\mm Y) = L\ln \frac{ \prod_{i = 1}^{\zeta_0} \phi_i}{\prod_{i = 1}^{\epsilon_0} \psi_i} 
        + L \left (\zeta_0- \epsilon_0  + \sum_{i = 1}^{\epsilon_0}\psi_i - \sum_{i = 1}^{\zeta_0}\phi_i\right ),
\end{align}
where $\{\psi_1, \cdots, \psi_{2M}\}$ are the eigenvalues of $\mm X/\sigma_r^2$ organized in decreasing order, $\{\phi_1, \cdots, \phi_{M}\}$ are the eigenvalues of $\mm X_d/\sigma_r^2$ organized in decreasing order, $\mm X = \mm Y \mm Y^H/L$ and $\mm X_d = \tilde{\mm Y}_d \tilde{\mm Y}_d^H/L$ are sample covariance matrices, and we have:
\begin{equation}
\label{eq:epsilon}
\epsilon_0 = \min\left( \sum_{i = 1}^{2M}\mm 1 (\psi_i \geq 1), C, 2M \right),
\end{equation}
\begin{equation}
\label{eq:zeta}
\zeta_0 = \min\left(\sum_{i = 1}^{M}\mm 1 (\phi_i \geq 1), C, M\right).
\end{equation}

\textit{Proof:} Please see Appendix A.

\section{Sensing Performance Analysis}\label{sec:Analysis}
To design the detection threshold and compute the detection probability, one must find the distribution of the test statistic. However, exact distributions of a complicated GLRT statistic are typically unavailable, except when the test statistic takes a special form \cite{santamaria2017passive}. Therefore, we resort to the asymptotic behavior of the detector (\ref{eq:statistic}) with a large sample length $L$ in this section.

\subsection{Asymptotic Distributions of the GLRT}

The hypothesis testing problem (\ref{eq:HT}) can be recast as a parameter test problem of the PDF (\ref{eq:PDF}):
\begin{equation}
\label{eq:paramtest}
    \begin{aligned} 
        \mathcal{H}_0: \quad \tilde{\mm H}_t = \mm 0, \tilde{\mm H}_d,\\
        \mathcal{H}_1: \quad \tilde{\mm H}_t \ne \mm 0, \tilde{\mm H}_d.\\
    \end{aligned}
\end{equation}

We define the set of all unknown parameters as $\bm \xi = [ \bm \xi_t^T, \bm \xi_d^T]^T \in \mathbb{C}^{4MC}$, where $\bm \xi_t= [\Re[\mstr{vec}(\tilde{\mm H}_t)]^T, \Im[\mstr{vec}(\tilde{\mm H}_t)]^T ]^T$ are the parameters of interest and $\bm \xi_d= [\Re[\mstr{vec}(\tilde{\mm H}_d)]^T, \Im[\mstr{vec}(\tilde{\mm H}_d)]^T ]^T$ are nuisance parameters. 
Under our signal model, it can be easily proved that the MLEs of these parameters can attain their asymptotic PDF (specific regularity conditions are described in \cite{kay1993fundamentalsestimation}). Let $\mm J(\bm \xi) \in \mathbb{C}^{2M\times 2M}$ denote the Fisher information matrix (FIM). It can be calculated as:
\begin{equation}
    \begin{aligned}
        \mathbf{J}(\bm \xi)
         = -\mathbb{E}\left[ \frac{\partial^2 \ln p(\mm Y | \mathcal{H}_j) }{\partial \bm \xi \partial \bm \xi^T}\right]
    \end{aligned}.
\end{equation} 

We partition the FIM into the following matrix:
\begin{equation}
\label{eq:fim1}
    \begin{aligned}
        \mathbf{J}(\bm \xi)
         = \begin{bmatrix} \mathbf{J}_{\bm \xi_t \bm \xi_t} & \mathbf{J}_{\bm \xi_t \bm \xi_d}\\ \mathbf{J}_{\bm \xi_d \bm \xi_t } & \mathbf{J}_{\bm \xi_d \bm \xi_d}\end{bmatrix}
    \end{aligned},
\end{equation} 
where $\mathbf{J}_{\bm \xi_t \bm \xi_t}, \mathbf{J}_{\bm \xi_t \bm \xi_d}, \mathbf{J}_{\bm \xi_d \bm \xi_t}, \mathbf{J}_{\bm \xi_d \bm \xi_d}\in \mathbb{C}^{M\times M}$. Then, the statistic $\Lambda(\mm Y)$ has the following distributions when $L \to \infty$ \cite{kay1993fundamentalsDetection}:
\begin{equation}
\label{eq:dist}
    \Lambda(\mm Y)  \overset{a}{\sim}
    \left \{ 
    \begin{aligned} 
        & \frac{1}{2} \chi^2(\nu),&\text{under }\mathcal{H}_0,\\
        & \frac{1}{2} {\chi'}^2(\nu, \kappa),&\text{under }\mathcal{H}_1,
    \end{aligned} 
    \right.
\end{equation}
where $\chi^2(\nu)$ represents chi-square distribution with degree-of-freedom $\nu$, and ${\chi'}^2(\nu, \kappa)$ represents non-central chi-square distribution with degree-of-freedom $\nu$ and non-centrality parameter $\kappa$, which is given by:
\begin{equation}
\label{eq:kappadef}
    \begin{aligned}
        \kappa = &(\bm \xi_{t, 1} - \bm \xi_{t, 0})^T\left (
    \mathbf{J}_{\bm \xi_{t, 0} \bm \xi_{t, 0} } -\mathbf{J}_{\bm \xi_{t, 0} \bm \xi_d} \mathbf{J}_{\bm \xi_d \bm \xi_d}^{-1} \mathbf{J}_{\bm \xi_d \bm \xi_{t, 0} } \right ) \\ & \quad (\bm \xi_{t, 1} - \bm \xi_{t, 0}),
    \end{aligned}
\end{equation}
where $\bm \xi_{t, 1}$ and $\bm \xi_{t, 0}$ are the true values of $\bm \xi_{t}$ under $\mathcal{H}_1$ and $\mathcal{H}_0$, respectively. Based on this, we give the asymptotic distribution of $\Lambda(\mm Y)$ in the following proposition.

\textit{Proposition 2:} For the asymptotic distributions of $\Lambda(\mm Y)$ at $L \to \infty$, the degree-of-freedom is given by $\nu = 2MC$, and the non-centrality parameter is given by:
\begin{equation}
\label{eq:kappa}
    \kappa = \frac{2L}{\sigma_r^2}\mtr\left [ \tilde{\mm H}_t \tilde{\mm H}_d^H (  \sigma_r^2\mm I_M + \tilde{\mm H}_d\tilde{\mm H}_d^H)^{-1} \tilde{\mm H}_d \tilde{\mm H}_t^H \right ].
\end{equation}

\textit{Proof:} Please see Appendix B.

Based on \textit{Proposition 2}, we can now obtain the asymptotic false alarm rate, denoted as $\tilde{P}_{\text{fa}}$, and the asymptotic detection probability, denoted as $\tilde{P}_{\text{d}}$, of the GLRT detector at $L \to \infty$. Specifically, for a decision threshold $\rho$, we have:
\begin{align}
    \label{eq:pfa}
    &\tilde{P}_{\text{fa}} = \mathbb{P}(\Lambda(\mm Y) > \rho; \mathcal{H}_0) = \mathcal F_{\chi^2(\nu)} (2\rho),\\
    \label{eq:pd}
    &\tilde{P}_{\text{d}} = \mathbb{P}(\Lambda(\mm Y) > \rho; \mathcal{H}_1) = \mathcal Q_{\frac{\nu}{2}}(\sqrt{\kappa}, \sqrt{2\rho}),
\end{align}
where $\mathcal Q_{x}(a, b)$ is the generalized Marcum Q-function with order $x$, and $\mathcal F_{\chi^2(\nu)}(\cdot)$ is the right-tail probability of the distribution $\chi^2(\nu)$.

\begin{figure}
\begin{centering}
\includegraphics[scale=0.45]{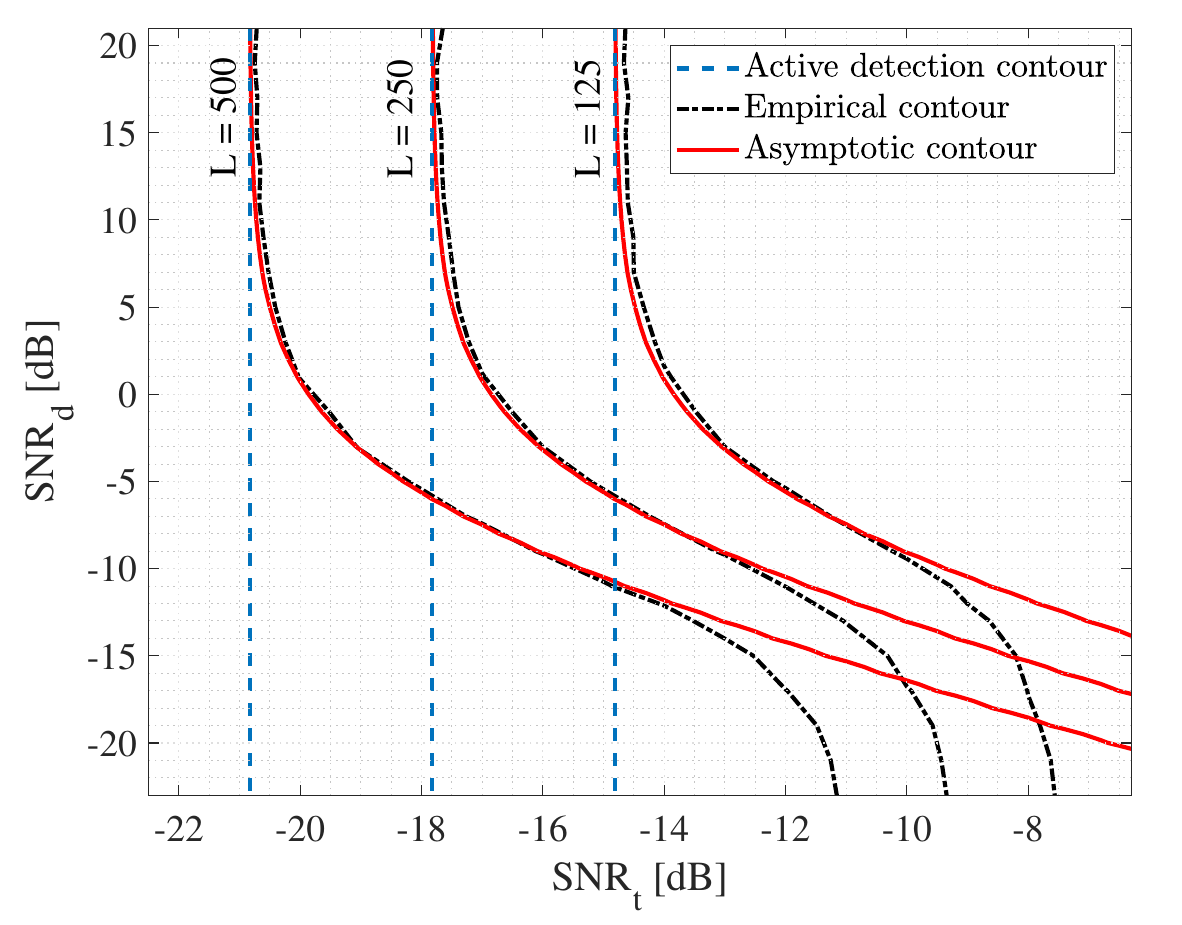}
\vspace{-0.6cm}
\caption{\small{$P_\text{d} = 0.9$ contour for the single-CU case where $C = 1$, $M = 4$, $P_\text{fa} = 10^{-3}$ and $L = [125, 250, 500]$.}}\label{fig:IsoContour}
\end{centering}
\vspace{-0.3cm}
\end{figure}

\subsection{Discussion} \label{sec:IIIB-discussion}
From (\ref{eq:pfa}) and (\ref{eq:pd}), it is known that while the asymptotic false alarm rate $\tilde{P}_{\text{fa}}$ only depends on the degree-of-freedom $\nu$, the asymptotic detection probability $\tilde{P}_{\text{d}}$ exhibits a monotonic increase with respect to the non-centrality parameter $\kappa$, which is a function of the sensing channel, the transmitted signal covariance $\mm R_{\text{c}}$, and the noise power $\sigma_r^2$, as shown in (\ref{eq:H}) and (\ref{eq:kappa}). This indicates that $\kappa$ can be interpreted as a composite measure of radar SINR in passive detection systems, and could be exploited as a performance indicator for system optimization. However, the intricate structure of $\kappa$ necessitates a more detailed analysis. To investigate the factors influencing detection performance and examine the accuracy of the asymptotic approximations relative to actual detectors, we first establish insights by looking into a simplified single-CU scenario, then extending the discussion to the general case.

\begin{figure}
\begin{centering}
\includegraphics[scale=0.45]{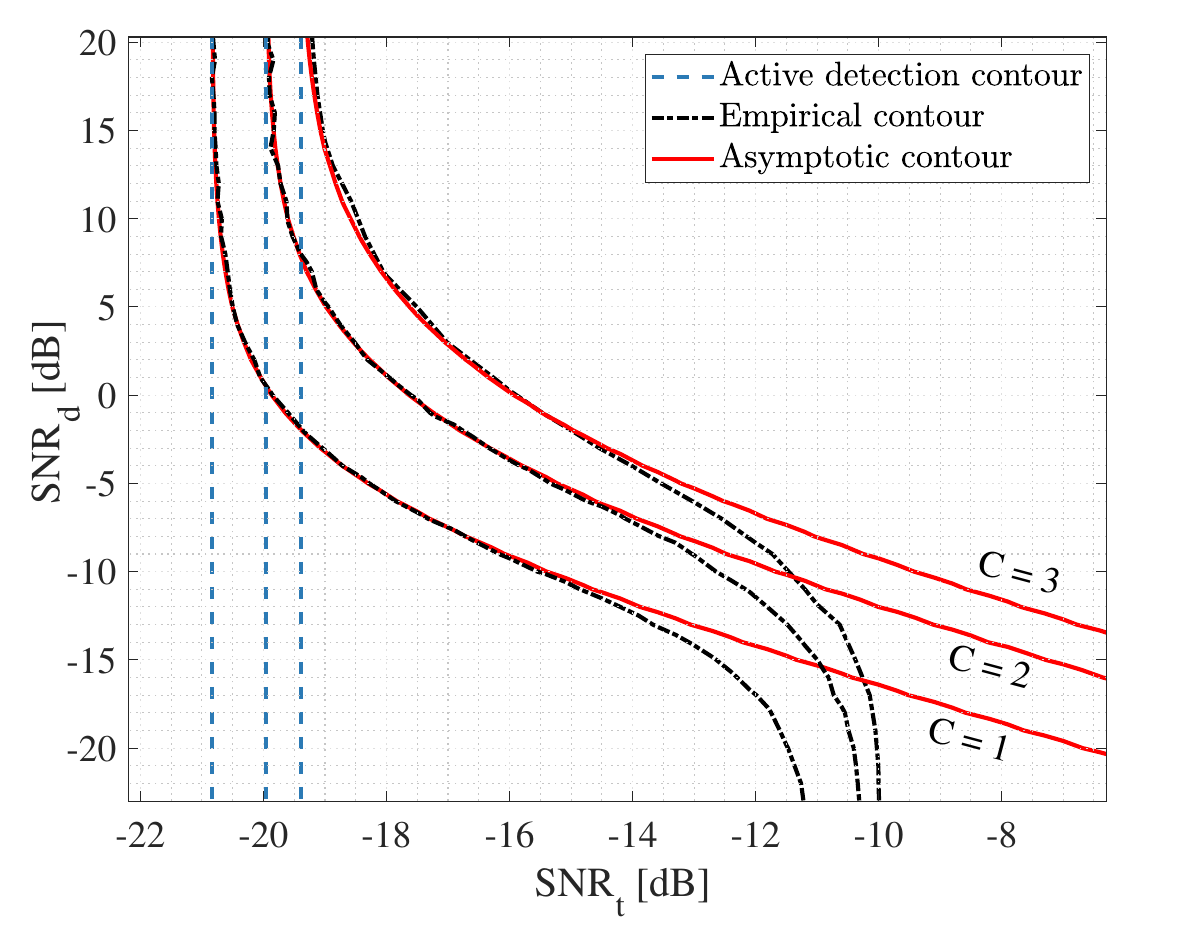}
\vspace{-0.6cm}
\caption{\small{$P_\text{d} = 0.9$ contour for multi-CU case where $L = 500$, $M = 4$, $P_\text{fa} = 10^{-3}$ and $C = [1, 2, 3]$.}}\label{fig:IsoContour2}
\end{centering}
\vspace{-0.3cm}
\end{figure}

\textit{1) Single-CU Case: }When there is only one CU, i.e., $C=1$, the equivalent channel matrices $\tilde{\mm H}_t$ and $\tilde{\mm H}_d$ reduce to vectors, and (\ref{eq:kappa}) can be simplified to the following formula:
\begin{equation}
\label{eq:kappa2}
    \kappa = 2LM^2 \text{SNR}_t\frac{\text{SNR}_d}{1+M\text{SNR}_d},
\end{equation}
where we have defined the average SNR of the $M$ target paths as:
\begin{equation}
\label{eq:snrt}
    \text{SNR}_t \triangleq \frac{1}{M\sigma_r^2} \mtr \left (\tilde{\mm H}_t \tilde{\mm H}_t^H \right),
\end{equation}
and the average SNR of the $M$ direct paths as:
\begin{equation}
\label{eq:snrd}
    \text{SNR}_d \triangleq \frac{1}{M\sigma_r^2} \mtr \left (\tilde{\mm H}_d \tilde{\mm H}_d^H \right).
\end{equation}

It can be observed that the asymptotic behavior of the GLRT statistic now only depends on the number of SRs $M$, the length of the data frame $L$, $\text{SNR}_t$, and ${\text{SNR}_d}$. Besides, for a fixed degree-of-freedom, we have $\tilde{P}_{\text{d}}$ increases monotonically with $\kappa$, while $\kappa$ increases monotonically with $\text{SNR}_t, \text{SNR}_d$, and $L$. This is consistent with the fact that the detection probability is higher with larger SNR in the target direction, better quality of the reference signal, and longer observation time. This observation shall motivate our second ISAC beamforming design problem as detailed in Section \ref{sec:BF} later. 

The dependencies of the detection probability on $\text{SNR}_t$ and $\text{SNR}_d$ in the single-CU case is shown in Fig. \ref{fig:IsoContour}, where the performance contours with $P_\text{d} = 0.9$ are depicted for the active detector derived in Appendix C, the empirical passive detector given in (\ref{eq:statistic}), and the asymptotic performance given by (\ref{eq:pfa})-(\ref{eq:kappa2}). Here, the active detector is derived in the same geometric configuration as the passive detector but has complete prior knowledge on the transmit signals. For each detector, the detection threshold is determined to achieve a false alarm rate of $10^{-3}$ based on $10^5$ Monte Carlo simulations. 

It can be observed that as $\text{SNR}_d$ increases, the asymptotic contour can better approximate the actual performance. As $L$ increases, this approximation becomes more accurate. In addition, when we have large $\text{SNR}_d$, the direct path provides a nearly noiseless reference for the SRs, and the corresponding detection probability approaches that of the active detection. The passive detection non-centrality parameter (\ref{eq:kappa2}) approaches $\kappa_{\text{act}} = 2LM \text{SNR}_t$ in active detection as derived in Appendix C. 

\textit{2) Multi-CU Case:} We first reveal that the asymptotic detection probability of the GLRT detector can still approach the performance upper bound obtained by active detection under some conditions.

\textit{Proposition 3:} The non-centrality parameter $(\ref{eq:kappa})$ can be expressed as:
\begin{equation}
\label{eq:refor_kappa}
    \kappa = \frac{2L}{\sigma_r^2} \sum_{n = 1}^C \frac{\bar{\sigma}_n}{1+\bar{\sigma}_n}\delta_n,
\end{equation}
where $\{\bar{\sigma}_1, \cdots, \bar{\sigma}_C\}$ are the eigenvalues of $\tilde{\mm H}_d^H\tilde{\mm H}_d/\sigma_r^2$ organized in decreasing order, $\delta_n = \mm v_n^H \tilde{\mm H}_t^H \tilde{\mm H}_t\mm v_n$ is the target energy on the direction of the $n$-th eigenvector $\mm v_n$ of the matrix $\tilde{\mm H}_d^H\tilde{\mm H}_d/\sigma_r^2$. When the minimum eigenvalue satisfies $\bar{\sigma}_C \gg 1$, we have $\kappa \approx 2LM \text{SNR}_t$.

\textit{Proof:} Please see Appendix D.

Note that the coefficient term $\frac{\bar{\sigma}_n}{1+\bar{\sigma}_n}$ in Equation (\ref{eq:refor_kappa}) indicates the performance loss of passive detection compared to the upper bound using active detection. Besides, the asymptotic passive detection probability could approach the upper bound when $\bar{\sigma}_C \gg 1$, which shows that a sufficient number of SRs and high SNRs of all direct paths are required. Moreover, the fractional structure in (\ref{eq:refor_kappa}) suggests that the target path terms $\{ \delta_n\}$ have a dominant influence on the passive detection performance, whereas the direct path terms $\left\{ \frac{\bar{\sigma}_n}{1+\bar{\sigma}_n}\right\}$ exert a constraining effect.

Next, we assess the derived distributions by comparing with empirical realizations. Equation (\ref{eq:kappa}) indicates that the asymptotic distribution of $\Lambda(\mm Y)$ under the alternative hypothesis is not solely dependent on aggregate parameters such as the average SNR, but rather on the specific channel realizations. To evaluate the accuracy of the asymptotic approximation to the actual performance, we present the performance contours with $P_\text{d} = 0.9$ for varying numbers of CUs in Fig. \ref{fig:IsoContour2}, where both the asymptotic and empirical contours are averaged over $10^5$ Monte Carlo simulations. It is evident that as the signal subspace dimension $C$ decreases, the detection performance improves, and the approximation exhibits better convergence to the empirical performance.

\section{Beamforming Optimization} \label{sec:BF}
In this section, we focus on the joint transmit beamforming design for downlink communication and passive detection based on previous analyses. Two different designs are proposed in the following subsections, namely, a ``$\max \tilde{P}_{\text{d}}$" design and a ``$\text{SNR}_d$ threshold" design.

\subsection{Beamforming Optimization by Maximizing \texorpdfstring{$\tilde{P}_{\text{d}}$}{Pd-tilde}}

In the preceding section, we demonstrated that the asymptotic result provides a reasonable approximation of the actual performance. Building on this, we now aim to optimize the beamformer by maximizing the asymptotic detection probability (\ref{eq:pd}), subject to constraints on the SINR for each of the $C$ CUs and a total transmit power budget. The resulting optimization problem can be formulated as follows:

\begin{subequations} \label{eq:op_original}
\begin{align}
    (\mathcal P1): & \quad \max_{\mm W} \quad \tilde{P}_{\text{d}} \tag{30}\\
    \label{eq:op_originalsub1}
    \text{s.t.} & \quad \sigma_{\text{c}}^2 + \mm h_n^H \mm R_{\text{c}} \mm h_n \leq (1+\Gamma_{c}^{-1})\mm h_n^H \mm R_n \mm h_n, \tag{30a} \\
    & \quad \forall n \in \{ 1, \cdots, C\}, \notag \\
    \label{eq:op_originalsub2}
    & \quad \mtr(\mm R_{\text{c}})\leq P_t, \tag{30b}
\end{align}
\end{subequations}
where $P_t$ is the power budget, $\mm R_n = \mm w_n \mm w_n^H$ and $\mm R_{\text{c}} = \sum_{n = 1}^C \mm R_n$ are transmit covariance matrices. Note that the false alarm probability does not depend on the joint beamforming design as discussed in Section \ref{sec:IIIB-discussion}, and therefore it is irrelevant to the above problem formulation. Furthermore, given the monotonically increasing relationship between $\tilde{P}_{\text{d}}$ and $\kappa$, problem $(\mathcal P1)$ is equivalent to maximizing $\kappa$. With (\ref{eq:H}) and (\ref{eq:kappa}), the objective can be rewritten as:
\begin{align}
    \kappa(\mm W)
    &= \frac{2L}{\sigma_r^2}\mtr\left [ \tilde{\mm H}_t^H\tilde{\mm H}_t \tilde{\mm H}_d^H (  \sigma_r^2\mm I_M + \tilde{\mm H}_d\tilde{\mm H}_d^H)^{-1} \tilde{\mm H}_d \right ] \notag\\
    &\triangleq 2L \mu_0 {\mm a}_{t}^H \mm R_{\text{c}} {\mm B}^H \left (  \mm I_M + {\mm B} \mm R_{\text{c}} {\mm B}^H \right)^{-1} {\mm B} \mm R_{\text{c}} {\mm a}_{t},
\end{align}
where we have denoted $\mu_0 \triangleq \sum_{i = 1}^M|\mu_{t, i}|^2/\sigma_r^2$, and:
\begin{equation}
    \mm B \triangleq \frac{1}{\sigma_r}\begin{bmatrix} 
    \mu_{d, 1} {\mm a}_{d, 1}^H  \\ \vdots
    \\ \mu_{d, M} {\mm a}_{d, M}^H  \end{bmatrix} \in \mathbb{C}^{M\times N_{t}}.
\end{equation}

We observe that the objective is highly non-convex with respect to $\mm W$ due to the multidimensional fractional structure. To handle this, we first introduce a quadratic transform with an auxiliary vector $\mm u$:
\begin{equation}
\label{eq:g}
    g(\mm W, \mm u) = 2\Re\{ \mm u^H\mm B\mm R_{\text{c}} {\mm a}_{t} \} - \mm u^H \left (  \mm I_M + {\mm B} \mm R_{\text{c}} {\mm B}^H \right) \mm u.
\end{equation}

As $ \mm I_M + {\mm B} \mm R_{\text{c}} {\mm B}^H \succ 0$, function $g(\mm W, \mm u)$ has a unique global maximum point $\mm u^{\star}$ when $\mm W$ is held fixed, which is given by:
\begin{equation}
\label{eq:ustar}
    \mm u^{\star} = \left (  \mm I_M + {\mm B} \mm R_{\text{c}} {\mm B}^H \right)^{-1}\mm B\mm R_{\text{c}}\mm a_t.
\end{equation}

Moreover, the optimal value at $\mm u^{\star}$ aligns with the objective for any $\mm W$:
\begin{equation}
\label{eq:Obalign}
    2L \mu_0 \max_{\mm u} g(\mm W, \mm u) = \kappa(\mm W).
\end{equation}

With this equality, we can reformulate the problem $(\mathcal P1)$ into the following equivalent form with $\mm W$ and $\mm u$:
\begin{subequations} \label{eq:op1.1}
\begin{align}
    (\mathcal P1.1): & \quad \max_{\mm W, \mm u} \quad g(\mm W, \mm u)\tag{36}\\
    \text{s.t.} & \quad (\ref{eq:op_originalsub1}), (\ref{eq:op_originalsub2}). \notag
\end{align}
\end{subequations}

This can be solved by employing an alternating optimization approach. For the first sub-problem with fixed $\mm W$, both constraints are independent of $\mm u$, and the optimal solution is given by (\ref{eq:ustar}). For the second sub-problem with fixed $\mm u$, the objective and the constraint (\ref{eq:op_originalsub1}) are still non-convex. Thus, we further apply the SDR technique by substituting the original variables $\{ \mm w_n \}_{n = 1}^C$ with rank-1 matrices $\{ \mm R_n \}_{n = 1}^C$. Then, we relax the rank-1 constraints. This leads to sub-optimal solutions for the original problem since the relaxation here is not guaranteed to be tight. But it simplifies the sub-problem into an SDP with linear objective and constraints:
\begin{subequations} \label{eq:subproblem2}
\begin{align}
    & (\mathcal P1.2): \notag \\
    & \max_{\{ \mm R_n \}_{n = 1}^C} \sum_{n = 1}^C\mtr\left[\left( 2\Re\{\mm a_t\mm u^H\mm B\} - \mm B^H\mm u \mm u^H \mm B \right) \mm R_n\right] - ||\mm u||^2 \tag{37}\\
    & \text{s.t.}  \quad (\ref{eq:op_originalsub1}), (\ref{eq:op_originalsub2}), \notag
\end{align}
\end{subequations}
which can be solved using convex optimization toolbox. Then, two sub-problems can be solved iteratively until convergence. Such convergence is guaranteed by the monotonicity and boundedness of objective values. Denote $\{ \mm R_n^{(k)} \}, \mm u_k$ as the solutions in the $k$-th iteration. Using (\ref{eq:g}) and (\ref{eq:Obalign}), a straightforward result is:
\begin{align}
    & \frac{1}{2L \mu_0 }\kappa(\mm W^{(k+1)}) = g(\mm W^{(k+1)}, \mm u^{(k+1)}) \geq  g(\mm W^{(k)}, \mm u^{(k+1)}) \notag \\
        & \quad \geq g(\mm W^{(k)}, \mm u^{(k)}) = \frac{1}{2L \mu_0 }\kappa(\mm W^{(k)}),
\end{align}
which suggests that the optimal value increases monotonically as iteration proceeds. Alongside with the limited transmit power constraint, the objective value will converge to a stationary point. The algorithm is summarized in \textit{Algorithm 1}. 

\begin{algorithm}
\caption{Alternating optimization algorithm for $(\mathcal P1)$}
\vspace{4pt}
\textbf{Input:} $\sigma_{\text{c}}^2, \sigma_r^2, \Gamma_c, \mm h_n, P_t, \mm a_t, \mm B, \epsilon, k_{max}$.

\textbf{Output:} Beamforming vectors $\{\mm w_n^*\}_{n=1}^C$.


\textbf{Steps:}
\begin{algorithmic}[1]
    \State Set $k = 0$, and initialize $\left\{ \mm R_n^{(0)} \right\}_{n= 1}^C$, $\mm u^{(0)}$ with feasible values.
    \State \textbf{repeat}
        \State \qquad Update $\mm u^{(k)}$ with (\ref{eq:ustar}).
        \State \qquad Update $\left\{ \mm R_n^{(k)} \right\}_{n= 1}^C$ with the optimal value of $(\mathcal P1.2)$.
        \State \qquad Update $k = k+1$.
    \State \textbf{until} the relative increase of the objective value is below a threshold $\epsilon$ or the number of iterations reaches $k_{max}$.
    \State Apply Gaussian randomization to obtain the rank-one solutions $\{\mm w_n^*\}_{n=1}^C$.
\end{algorithmic}
\end{algorithm}

\subsection{Beamforming Optimization with Thresholds on \texorpdfstring{$\text{SNR}_d$}{SNRd}}
Iteratively solving the problem (\ref{eq:op1.1}) leads to high complexity. This motivates us to explore a simpler beamforming design. In Section \ref{sec:Analysis}, we have shown that the passive detection performance is improved with larger $\text{SNR}_t$ and $\text{SNR}_d$. Empirically, considering the dominant role of $\text{SNR}_t$ in (\ref{eq:refor_kappa}), we can design the transmit beamforming to maximize the energy in the target direction while imposing a minimum threshold on $\text{SNR}_d$. This threshold is used to preserve enough energy in the reference channel for the passive detector to deal with the unknown transmitted signals.

Based on the above discussion, we can formulate the second beamforming design problem as:
\begin{subequations} \label{eq:opt2}
\begin{align}
\label{eq:emobj}
    (\mathcal P2): & \quad \max_{\mm W} \quad {\mm a}_{t}^H \mm R_{\text{c}} {\mm a}_{t}  \tag{39}\\
    \label{eq:emc1}
    \text{s.t.} & \quad \text{SNR}_d \geq \Gamma_d \tag{39a}, \\
    & \quad (\ref{eq:op_originalsub1}), (\ref{eq:op_originalsub2}), \notag
\end{align}
\end{subequations}
which can be readily solved using the standard SDR and randomization technique. 

\textit{Remark:} Though such an empirical design is much simpler than solving $(\mathcal P1)$, the objective (\ref{eq:emobj}) and constraint (\ref{eq:emc1}) now lose the fractional structure in (\ref{eq:kappa}) and thereby cause performance degradation. Besides, tradeoff between the target paths and direct paths now has to be controlled manually by setting an threshold $\Gamma_d$. If the threshold is set too low, detection performance is mainly limited by the direct path quality; conversely, if the threshold is set too high, excessive energy is assigned for the direct paths, and the detection performance is predominantly constrained by the target path energy. However, if $\Gamma_d$ is appropriately set (for example, by searching within a finite interval), the solution from $(\mathcal P2)$ is equivalent to a relaxed version of $(\mathcal P1)$, as shown in Appendix E. Detection performance with varying choices of $\Gamma_d$ is evaluated numerically in Section \ref{sec:simulation}.

\subsection{Complexity Analysis}
For the ``$\max \tilde{P}_{\text{d}}$" design in \textit{Algorithm 1}, the computation complexity is dominated by solving $(\mathcal P1.2)$. The worst case complexity with the primal-dual interior-point algorithm is estimated as $\mathcal{O}(TN_t^4 \log (1/\epsilon))$, where $T$ and $\epsilon$ are the alternating iteration number and solution accuracy, respectively. For the ``$\text{SNR}_d$ threshold" design, the complexity is estimated as $\mathcal{O}(N_t^4\log (1/\epsilon))$.

\section{Simulation Results} \label{sec:simulation}
In this section, we present numerical results to evaluate the proposed ``$\max \tilde{P}_{\text{d}}$" and ``$\text{SNR}_d$ threshold" beamforming designs.

\subsection{Simulation Setup}
\textbf{1) Simulation Parameters: }
In the following simulations, unless otherwise specified, we assume that the BS has $N_t = 16$ antennas, and each SR has $N_{1} = 14$ antennas for the surveillance array and $N_{2} = 2$ antennas for the reference array. For the active detection scheme, we use $N_{r} = 14$ receiving antennas. The transmit power is $-10$ dBW. The noise power of communication and sensing is set as $\sigma_r^2 = \sigma_{\text{c}}^2 = -114$ dBW. The target reflection coefficient is assumed to follow the complex Gaussian distribution $\mathcal{CN}(0, \sigma_{RCS}^2)$ and $\sigma_{RCS}^2 = 0$ dBsm corresponds to the average target RCS. The false alarm rate is set as $10^{-3}$. The data frame length is $L = 500$.  The BS is located at the coordinate $(0 \text{m}, 0 \text{m})$. There are $M=4$ SRs located at the coordinates $(141.4 \text{m}, 141.4 \text{m})$, $(141.4 \text{m}, -141.4 \text{m})$, $(-141.4 \text{m}, 141.4 \text{m})$ and $(-141.4 \text{m}, -141.4 \text{m})$. A target is placed at $(0 \text{m}, -100 \text{m})$. Besides, there are $C = 2$ CUs placed $100$ m away from the BS. Rayleigh fading is assumed for the communication channel. The propagation loss of the target and direct paths are calculated by \cite{5776640}:
\begin{equation}
    \begin{aligned}
        \mathrm{PL}_{t, i}&=\frac{\lambda^2}{(4\pi )^3||\mm p_{\text{SR}, i}-\mm p_{\text{tar}}||^2 \cdot||\mm p_{\text{BS}}-\mm p_{\text{tar}}||^2}, \\
        \mathrm{PL}_{d, i}&=\frac{\lambda^2}{(4\pi )^2||\mm p_{\text{SR}, i}-\mm p_{\text{BS}}||^2}.
    \end{aligned}
\end{equation}

\textbf{2) Benchmark Schemes}:
For comparison, we consider four benchmark schemes detailed as follows:
\begin{itemize}
    \item \textit{Active detection.} This scheme serves as a performance upper bound for passive detection, where the SRs have perfect knowledge of the transmitted signals and perform active detection (the active detector is derived in Appendix C). The BS and SRs are in the same multi-static geometry as our proposed schemes. The beamformer can be obtained by solving:
    \begin{subequations} \label{eq:activeBF}
    \label{eq:28}
    \begin{align}
        \max_{\mm W} & \quad {\mm a}_{t}^H \mm R_{\text{c}} {\mm a}_{t} \tag{41} \\
        \text{s.t.} & \quad (\ref{eq:op_originalsub1}), (\ref{eq:op_originalsub2}),  \notag
    \end{align}
    \end{subequations}
    which can be directly solved using the standard SDR technique. 

    \item \textit{Max $\text{SNR}_t$ w/o $\text{SNR}_d$ threshold.} This scheme utilizes a conventional beamforming method for passive detection. Here, the proposed detector (\ref{eq:statistic}) is used for detection. But the beamformer is obtained by solving (\ref{eq:activeBF}). This means that the quality of direct paths is not guaranteed.

    \item \textit{Communication-only.} This scheme examines passive detection performance when the BS is not collaborative for sensing. The beamforming matrix is obtained solving the feasibility problem of $(\mathcal P1)$, which corresponds to any feasible solution satisfying the communication and power constraints. Under this scenario, the detection system completely relies on opportunistic illumination, and is essentially a passive radar system.

    \item \textit{Sensing-only.} This scheme provides another upper bound for the passive detection performance without communication-sensing tradeoff. The beamforming problems $(\mathcal P1.2)$ and $(\mathcal P2)$ are solved without communication constraints.
\end{itemize}

\subsection{Comparison of Beamforming Designs}

\begin{figure}
\begin{centering}
\includegraphics[scale=0.43]{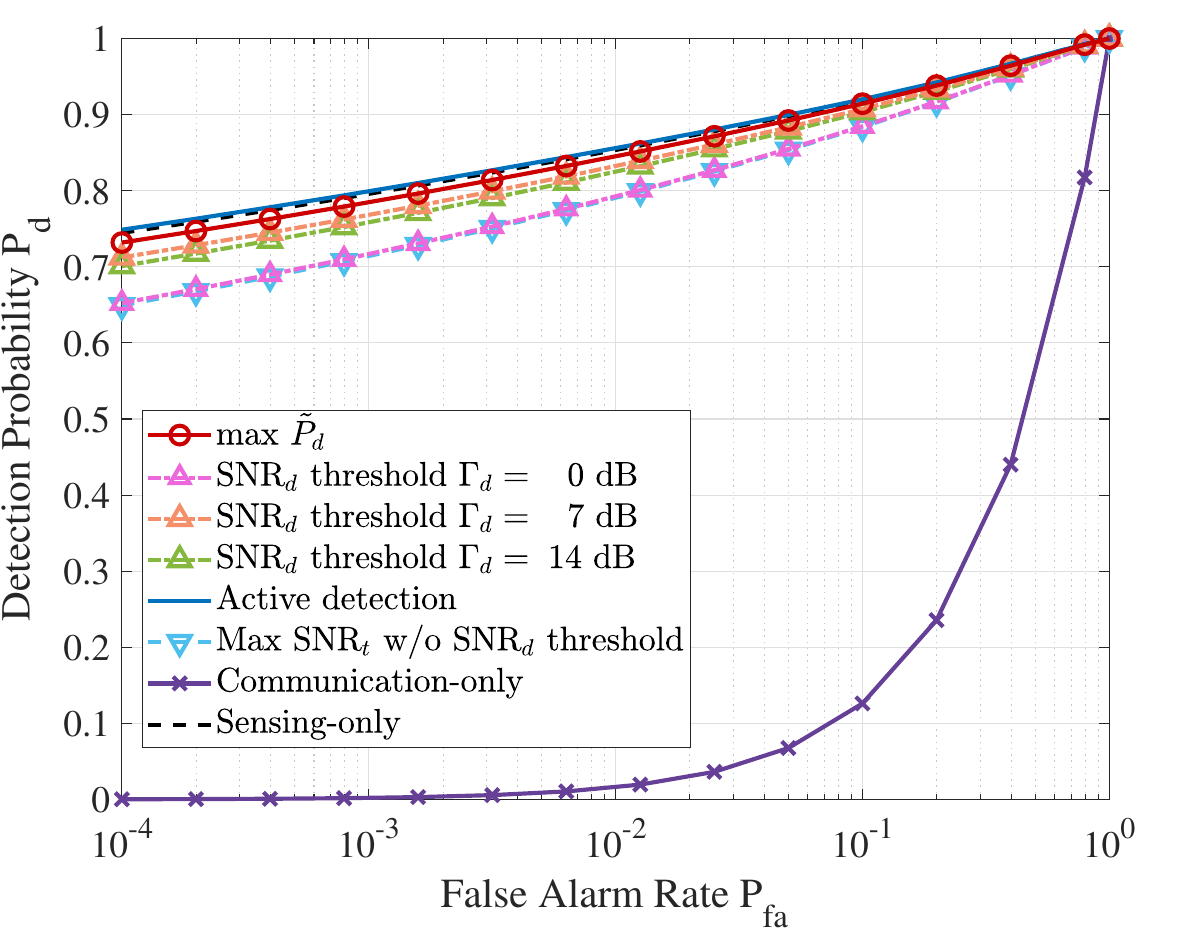}
\caption{\small{The receiver operating characteristic with $C = 2$ CUs, $M = 4$ SRs, and $\Gamma_{\text{c}} = 12$ dB.}}\label{fig:ROC}
\end{centering}
\end{figure}
\begin{figure}
\begin{centering}
\includegraphics[scale=0.43]{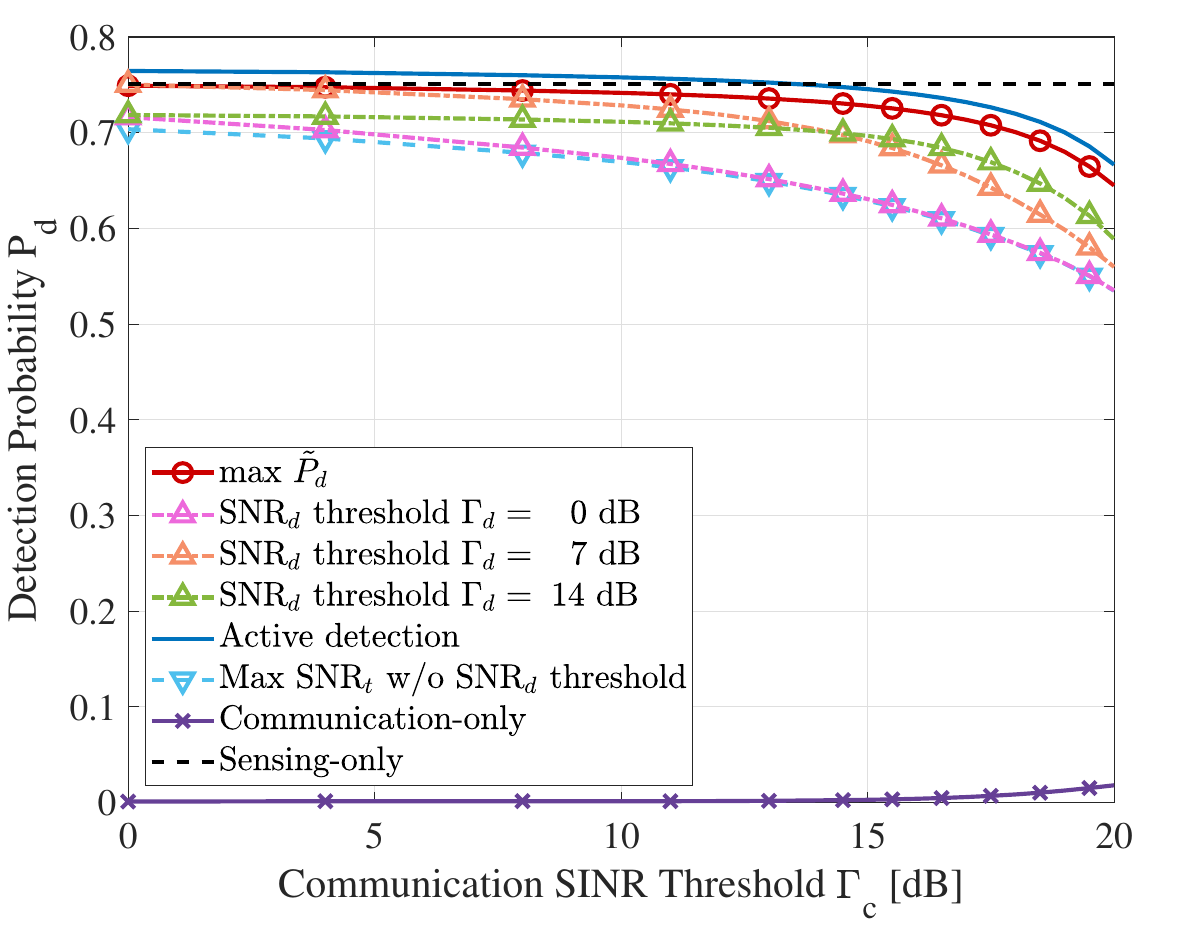}
\caption{\small{Detection probability versus the communication SINR requirement $\Gamma_{\text{c}}$.}}\label{fig:Tradeoff1}
\end{centering}
\end{figure}

In this section, we analyze the communication and sensing performance of proposed designs and benchmark schemes under various system parameters.

\textit{1) Target detection performance:} 
We start by assessing the detection performance using receiver operating characteristic (ROC) curves. Figure \ref{fig:ROC} illustrates the probability of detection versus the false alarm rate under a communication SINR threshold of $\Gamma_{\text{c}} = 12$ dB. The results indicate that the ``$\max \tilde{P}_{\text{d}}$" design achieves performance closer to the upper bounds than the other designs. For the ``$\text{SNR}_d$ threshold" design, detection performance is closely related to the pre-set thresholds. When an appropriate value is selected, its performance can approach that of the ``$\max \tilde{P}_{\text{d}}$" design. Furthermore, detection performance undergoes significant degradation when the BS operates with communication-only beamforming. This is due to the generally low average opportunistic illumination power directed towards both the target and reference arrays, as highly directive beams are only focused on the CUs. However, if the BS collaborates in beamforming at the target using a conventional active detection beamformer, performance improves substantially. This improvement can be attributed to the dominant influence of $\text{SNR}_t$ on performance, as compared to $\text{SNR}_d$, as discussed in section \ref{sec:Analysis}.
\begin{figure}
\begin{centering}
\includegraphics[scale=0.43]{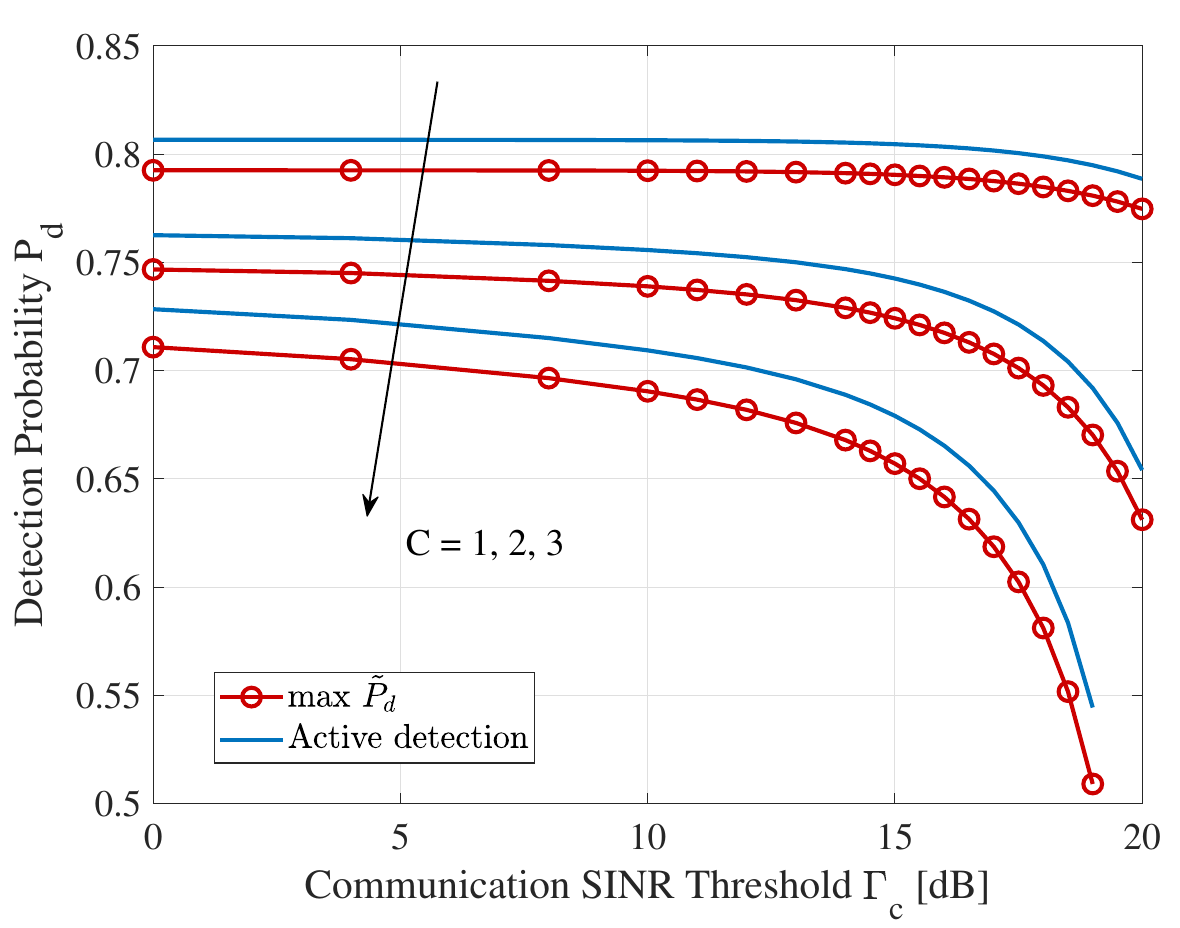}
\caption{\small{Trade-off curves with varying numbers of CUs. Here $C = [1, 2, 3]$, $P_{\text{fa}} = 10^{-3}$, $L = 500$.}}\label{fig:numC}
\end{centering}
\end{figure}

\textit{2) Tradeoff between communication and sensing:} 
In Fig. \ref{fig:Tradeoff1}, we investigate the interplay between communication and sensing in terms of the tradeoff between detection probability and communication SINR threshold. As expected, the detection probability decreases as the SINR requirement becomes more stringent. The proposed ``$\max \tilde{P}_{\text{d}}$" design consistently outperforms the other design approaches. For the ``$\text{SNR}_d$ threshold" design, marginal improvements are observed when the threshold is set too low. With an appropriately chosen threshold, the performance is comparable to that of the ``$\max \tilde{P}_{\text{d}}$" design. In addition, passive detection is ineffective in the absence of collaborative beamforming from the BS. 

\begin{figure}
\begin{centering}
\includegraphics[scale=0.42]{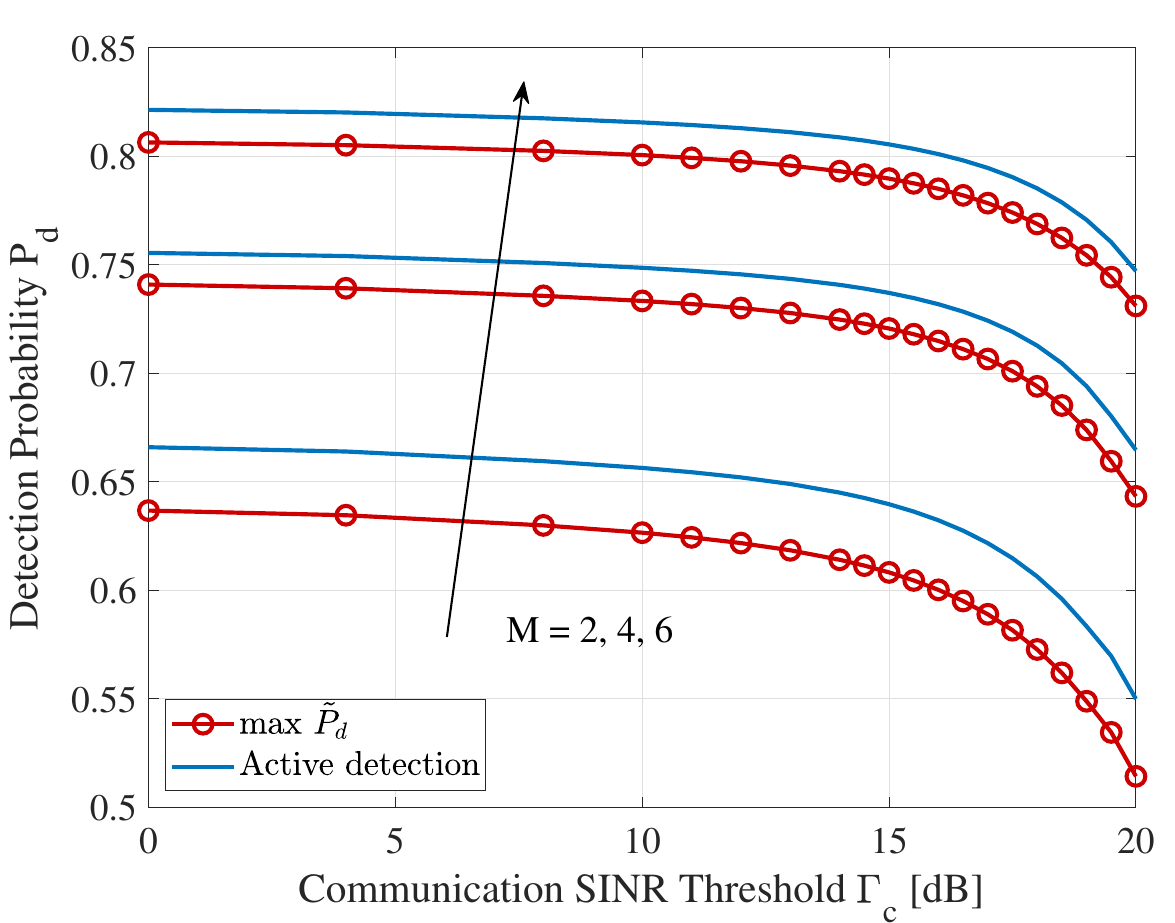}
\caption{\small{Trade-off curves with varying numbers of SRs. Here $M = [2, 4, 6]$, $P_{\text{fa}} = 10^{-3}$, $L = 500$.}}\label{fig:Nr}
\end{centering}
\end{figure}

\begin{figure}
\begin{centering}
\includegraphics[scale=0.42]{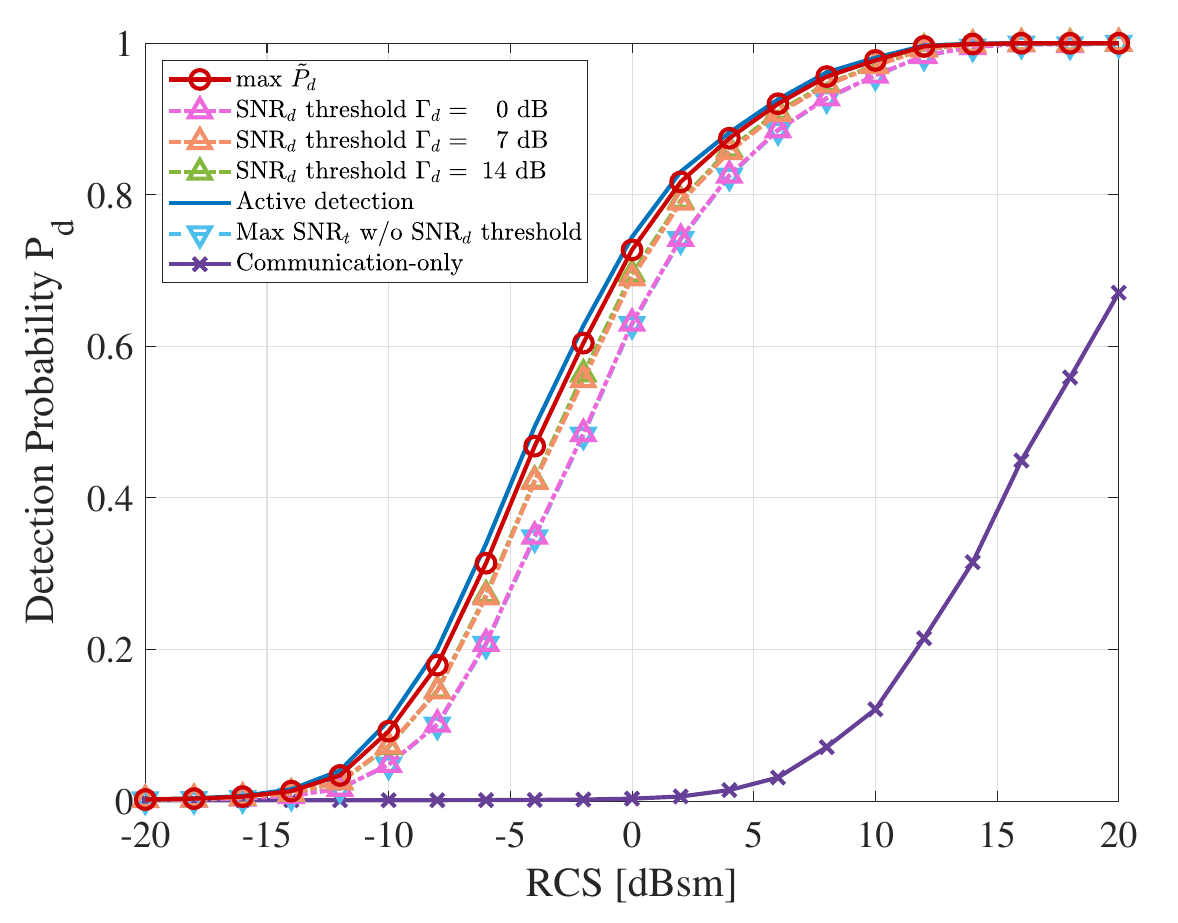}
\caption{\small{Detection probability versus the target RCS. Here $M = 4$, $C = 2$, $\Gamma_{\text{c}} = 12$ dB, $P_{\text{fa}} = 10^{-3}$, $L = 500$.}}\label{fig:RCS}
\end{centering}
\end{figure}

\textit{3) Impact of the number of CUs and SRs:} 
In Fig. \ref{fig:numC}, we present the tradeoff curves for varying numbers of CUs. It is observed that $P_{\text{d}}$ decreases as the number of CUs increases for both schemes. Besides, the performance gap between active and passive detection widens with the increasing number of CUs. These results are consistent with the analyses presented in Section \ref{sec:Analysis}. 

In Fig. \ref{fig:Nr}, we evaluate the detection performance for varying numbers of SRs. As anticipated, detection performance improves with an increasing number of SRs as the total energy received from the target increases. In addition, the performance gap between active and passive detection narrows as the number of SRs increases.

\begin{figure}
\begin{centering}
\includegraphics[scale=0.42]{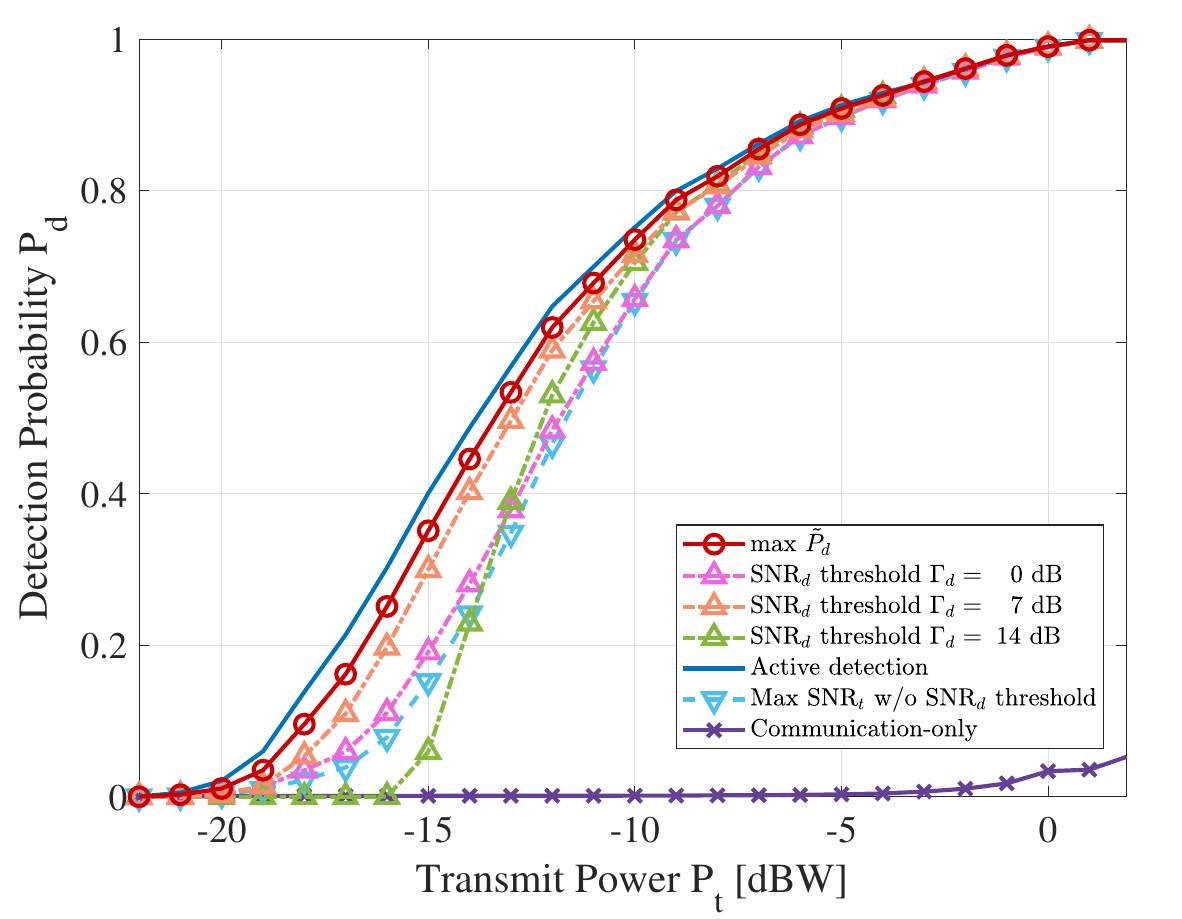}
\caption{\small{Detection probability versus the transmit power. Here $M = 4$, $C = 2$, $\Gamma_{\text{c}} = 12$ dB, $P_{\text{fa}} = 10^{-3}$, $L = 500$.}}\label{fig:Pt}
\end{centering}
\end{figure}

\textit{4) Impact of RCS and transmit power:} 
In Fig. \ref{fig:RCS}, we evaluate the detection performance for varying target RCS strengths. Larger RCS implies better detection performance for all design approaches as expected. The performance gap between different designs diminishes as the RCS increases. Fig. \ref{fig:Pt} illustrates the detection probability under different power budget, where higher power leads to better detection performance and the ``$\max \tilde{P}_{\text{d}}$" design demonstrates an evident improvement over other designs. 

\subsection{Detection with OFDM signals}
In Section \ref{sec:Analysis} and the above simulations, the transmit signal is assumed to follow Gaussian distribution to keep the analyses trackable. To further validate the performance of passive detection and the proposed beamformming design in actual ISAC systems, we now perform simulations using orthogonal frequency division multiplexing (OFDM) signals. The sensing procedure is illustrated in Section \ref{sec: II-B}. Specifically, the delay-Doppler domain is discretized into $N_0=10000$ grid cells. We use OFDM signals with $N_{sc} = 1024$ subcarriers. The subcarrier spacing is $\Delta_f = 30$ kHz. We assume that the BS provides downlink communication to one CU with a distance of $150$ m from the BS and an angle of $\theta=-60^\degree$. The user data is randomly generated and modulated using 16-QAM. Here, we adopt the Saleh-Valenzuela model for the user channel. Besides, there are $M=2$ SRs located at the coordinates $(225 \text{m}, 129.9 \text{m})$ and $(225 \text{m}, -129.9 \text{m})$. The BS is located at $(0 \text{m}, 0 \text{m})$. A target is placed at $(150 \text{m}, 0 \text{m})$.

\begin{figure}
\begin{centering}
\includegraphics[scale=0.40]{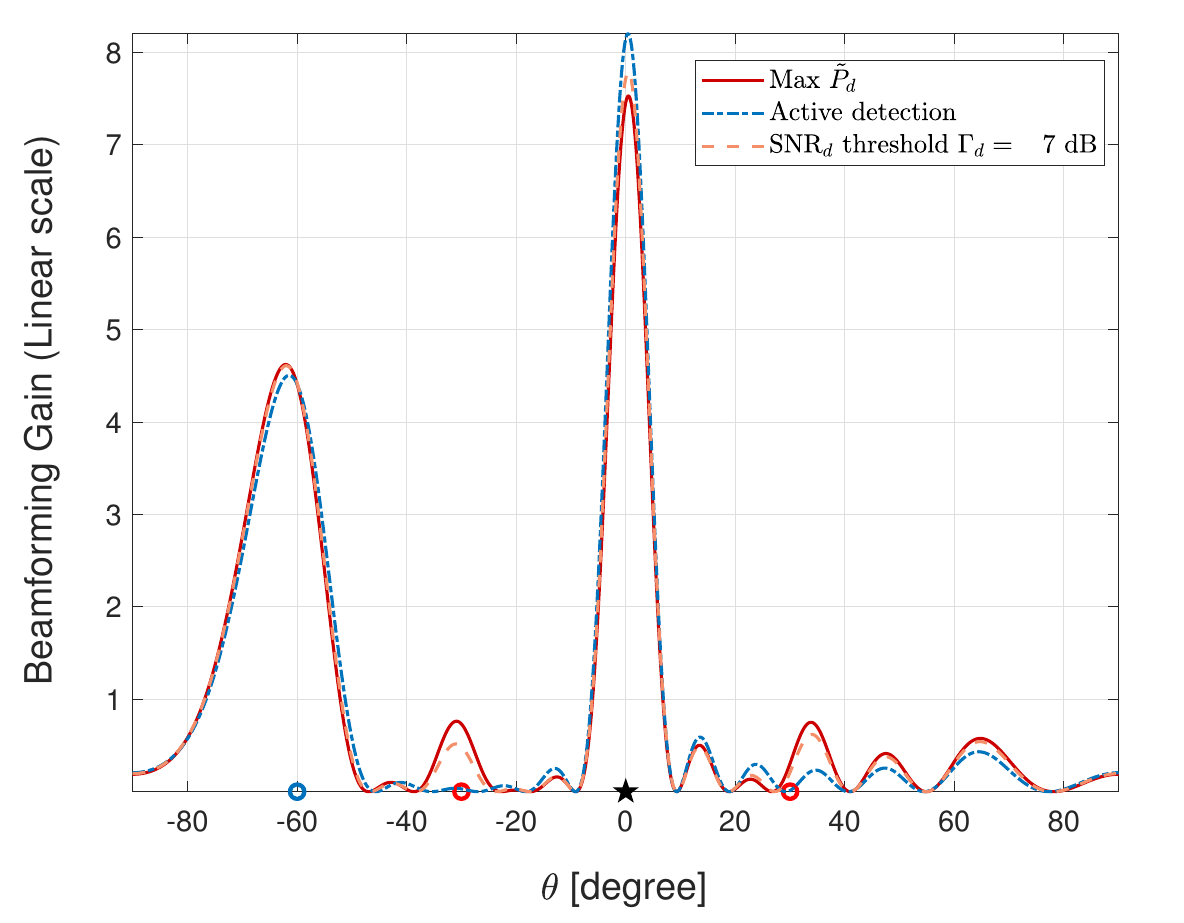}
\caption{\small{Beampatterns of different beamforming designs. The red circle, blue circle and black star refer to the directions of the SRs, CU and target, respectively.}}\label{fig:beam}
\end{centering}
\vspace{-0.3cm}
\end{figure}

In Figure \ref{fig:beam}, we analyze the transmit beampatterns of three distinct designs. The results demonstrate that, unlike active detection beamforming, the proposed design must simultaneously direct energy toward both the target path and direct paths. However, this necessitates sacrificing some energy in the target mainlobe, which consequently leads to performance degradation. Therefore, the proposed designs entail an ability to balance the competing demands of the target and direct paths. For the ``$\text{SNR}_d$ threshold" design, setting the threshold too low compromises the quality of the direct path, whereas an excessively high threshold results in inefficient energy allocation. In contrast, the ``$\max \tilde{P}_{\text{d}}$" design exhibits a superior ability to handle the tradeoff between the target and direct paths, thereby achieving enhanced overall performance.

\begin{figure}
\begin{centering}
\includegraphics[scale=0.39]{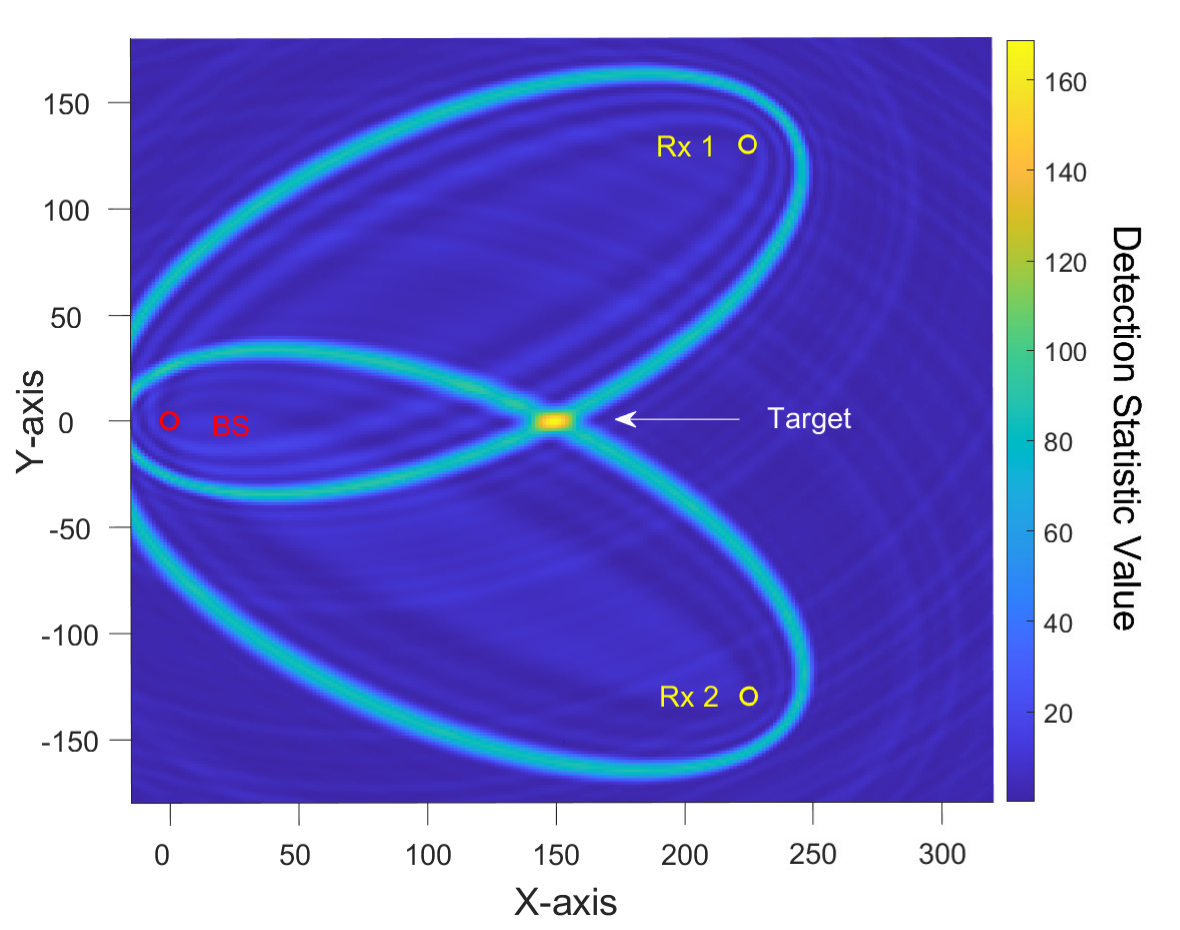}
\caption{\small{Detection heatmap.}}\label{fig:map}
\end{centering}
\vspace{-0.3cm}
\end{figure}

Using the optimized beam, we further illustrate the detection heatmap in Fig. \ref{fig:map}, where the horizontal and vertical axes represent spatial coordinates, with the heat value indicating the detection statistic calculated by (\ref{eq:statistic}). In particular, a distinct peak emerges within the grid corresponding to the true location of the target. Additionally, two elliptical sidelobes are evident, with the BS and SRs positioned as the foci. This pattern arises due to the cross-correlation operation of the detector, which captures the property that points along each ellipse share the same time delay as the target.

\section{Conclusion}
In this paper, we have investigated the passive detection problem in multi-static ISAC systems, focusing on the use of random unknown communication signals for target detection and the joint beamforming design. By deriving a GLRT detector and conducting detection performance analyses under the large-sample regime, we have revealed that the SNR of target paths plays a dominant role in determining the detection probability, while the SNR of direct paths excerts a constraining effect. Building on the analysis, we proposed two joint transmit beamforming designs. The first design maximizes the asymptotic detection probability while satisfying the power and communication SINR constraints. Despite the non-convex nature of this optimization problem, we developed an effective alternating optimization algorithm leveraging quadratic transform and semi-definite relaxation techniques. The second beamforming design maximizes the target energy while guaranteeing the worst direct path SNR, offering a computationally efficient alternative. The simulation results demonstrated the effectiveness of our proposed designs and ascertained the possibility of achieving reasonable passive detection performance in an ISAC system using unknown communication data signals.

{\section*{Appendix}
\subsection{Proof of Proposition 1} \label{detector}
Under hypothesis $\mathcal{H}_1$, the log-likelihood is given by:
\begin{equation}
    l_1 = -L\ln \det \mm R_{y, 1} - L \mathrm{tr} (\mm{R}_{y, 1}^{-1} \mm X),
\end{equation}
where $\mm R_{y,1} = \tilde{\mv H} \tilde{\mv H}^H + \sigma_r^2 \mm{I}_{2M}$, $\tilde{\mv H} = [\tilde{\mv H}_t^T, \tilde{\mv H}_d^T]^T$, and the constant term is omitted for brevity. We first calculate the spectral decomposition of $\tilde{\mm X}$:
\begin{equation}
    \tilde{\mm X} = \frac{1}{\sigma_r^2} \mm X = \Gamma \Psi \Gamma^H,
\end{equation}
where $\Psi = \mathrm{diag} ([\psi_1, \cdots, \psi_{2M}])$ contains the eigenvalues organized in decreasing order. Then, we utilize the theorem \textit{9.4.1} in \cite{mardia2024multivariate} to obtain the MLE of $\mm R_{y, 1}$:
\begin{equation}
    \hat{\mm R}_{y, 1} = \sigma_r^2 \mm I_{2M} + \sigma_r^2 {\Gamma} \mm B {\Gamma}^H,
\end{equation}
where $\mm B = \mathrm{diag} ([b_1, \cdots, b_m, 0, \cdots, 0]) \in \mathbb{C}^{ 2M \times 2M}$, $b_i = \max(\psi_i-1, 0)$, and $m=\min(C, 2M)$. We then calculate the maximized log-likelihood function:
\begin{align}
    \det(\hat{\mm R}_{y, 1}^{-1} \mm X) 
    &= \det \left((\mm I_{2M} + {\Gamma} \mm B  {\Gamma}^H )^{-1}\tilde{\mm X} \right ) \notag \\
    &= \det \left({\Gamma}(\mm I_{2M} + \mm B)^{-1} \Psi {\Gamma} \right ) \notag \\
    &= \prod_{i = 1}^{C}\min (\psi_i , 1) \prod_{i = C+1}^{2M} \psi_i,
\end{align}
\begin{equation}
    \mathrm{tr}(\hat{\mm R}_{y, 1}^{-1} \mm X) = \sum_{i = 1}^{C}\min (\psi_i , 1) + \sum_{i = C+1}^{2M} \psi_i,
\end{equation}
\begin{equation}
    \hat{l}_1 = -L\ln \det \mm X +L\ln \det(\hat{\mm R}_{y, 1}^{-1} \mm X) - L \mathrm{tr}(\hat{\mm R}_{y, 1}^{-1} \mm X).
\end{equation}

Under hypothesis $\mathcal{H}_0$, the derivation is similar, which gives:
\begin{align}
    \hat{l}_0 &= - LM\ln \sigma_r^2 + L\ln \det (\hat{\mm R}_{0}^{-1}\mm X_d) - L\ln \det \mm X_d \notag \\
    &\qquad - L\mathrm{tr} (\hat{\mm R}_{0}^{-1}\mm X_d) - \frac{L}{\sigma_r^2}\mathrm{tr} (\mm X_t),
\end{align}
\begin{equation}
    {\hat{\mm R}}_{0} = \sigma_r^2 \mm I_{M} + \sigma_r^2 {\Gamma_0} \mm B_0  {\Gamma^H_0},
\end{equation}
\begin{equation}
    \tilde{\mm X}_d = \frac{1}{\sigma_r^2} \mm X_d = \Gamma_0 \Phi \Gamma_0^H,
\end{equation}
where $\mm B_0 = \mathrm{diag} ([b_{0, 1}, \cdots, b_{0, m_0}, 0, \cdots, 0]) \in \mathbb{C}^{ M \times M}$, $b_{0, i} = \max(\phi_i-1, 0)$, $m_0 = \min(C, M)$, $\mm X_t = \tilde{\mm Y}_t\tilde{\mm Y}_t^H/L$, $\Phi = \mathrm{diag} ([\phi_1, \cdots, \phi_{M}])$ contains the eigenvalues in decreasing order. Finally, the GLRT statistic can be calculated:
\begin{equation}
    \begin{aligned}
        \Lambda(\mm Y) &= \hat{l}_1 - \hat{l}_0\\
        &= L\ln \frac{ \prod_{i = 1}^{\zeta_0} \phi_i}{\prod_{i = 1}^{\epsilon_0} \psi_i} 
        + L (\zeta_0- \epsilon_0  + \sum_{i = 1}^{\epsilon_0}\psi_i - \sum_{i = 1}^{\zeta_0}\phi_i ),
    \end{aligned}
\end{equation}
where $\epsilon_0$ and $\zeta_0$ are given by (\ref{eq:epsilon}) and (\ref{eq:zeta}).

\subsection{Proof of Proposition 2} \label{app}

Referring to \cite{kay1993fundamentalsDetection}, the degree-of-freedom $\nu$ of the asymptotic distribution is given by the dimension of the parameter of interest $\bm \xi_t$, which is $2MC$ here. 

To calculate $ \kappa$, we denote $\mm Y = \mm H \mm S + \mm Z$, where $\mm H = \mm C_1 \tilde{\mv H}_t+ \mm C_2 \tilde{\mv H}_d$ , $\mm C_1 = [\mm I_M, \mm 0_M]^T$, $\mm C_2 = [\mm 0_M, \mm I_M]^T$. Here we first calculate the FIM w.r.t. $\mstr{vec}(\mm H)$, then apply the chain rule to obtain the FIM w.r.t. $\bm \xi$. The log-likelihood is given by:
\begin{align}
    l &= - 2LM\ln \pi -L \ln \left ( \mstr{det}\mm R_{y, j} \right ) - L\mtr (\mm X \mm R_{y, j}^{-1}).
\end{align}

The first-order differential is then given by:
\begin{equation}
    \begin{aligned}
        \mstr{d} l &= \mtr(\mm F_1 \mstr{d} \mm H + \mm F_1^* \mstr{d} \mm H^*),\\
        \mm F_1 &= L(\mm H^H \mm R_{y, j}^{-1}\mm X \mm R_{y, j}^{-1} - \mm H^H \mm R_{y, j}^{-1}).
    \end{aligned}
\end{equation}

Taking the second-order differential, the full Hessian matrix is given by:
\begin{equation}
    \begin{aligned}
    &\mathcal{H} = \begin{bmatrix}
        \mathcal{H}_{\mm H, \mm H^*}& \mathcal{H}_{\mm H^*, \mm H^*}\\
        \mathcal{H}_{\mm H, \mm H}& \mathcal{H}_{\mm H^*, \mm H}
    \end{bmatrix},
    \\
     &\mathcal{H}_{\mm H, \mm H}= \mathcal{H}_{\mm H^*, \mm H^*}^* = \frac{1}{2}L\mm K\sum_{i = 1}^3 (\mm V_i^T \otimes \mm U_i + \mm U_i^T \otimes \mm V_i),
    \\
     &\mathcal{H}_{\mm H, \mm H^*}= \mathcal{H}_{\mm H^*, \mm H}^* = \frac{1}{2}L \sum_{i = 1}^3 (\bar{\mm V}_i^T \otimes \bar{\mm U}_i + \bar{\mm U}_i^T \otimes \bar{\mm V}_i),
    \end{aligned}
\end{equation}
where $\mm K$ is a commutation matrix, and:
\begin{equation}
    \left\{ \quad
    \begin{aligned}
    &\mm U_1 = \mm V_1 = -\mm U_2 = \mm V_3 = \mm H^H \mm R_{y, j}^{-1},\\
    &\mm V_2 = -\mm U_3 = \mm H^H \mm R_{y, j}^{-1}\mm X \mm R_{y, j}^{-1},\\
    &\bar{\mm U}_1 = -\bar{\mm U}_2= \mm H^H \mm R_{y, j}^{-1}\mm H - \mm I_C,\\
    &\bar{\mm V}_1 = \bar{\mm V}_3 = \mm R_{y, j}^{-1},\\
    &\bar{\mm V}_2 = \mm R_{y, j}^{-1}\mm X \mm R_{y, j}^{-1},\\
    &\bar{\mm U}_3 = - \mm H^H\mm R_{y, j}^{-1}\mm X \mm R_{y, j}^{-1}\mm H.
    \end{aligned}
    \right.
\end{equation}

The expectation is calculated as:
\begin{equation}
    \begin{aligned}
    &\mm {J}(\mm H, \mm H^*) = \begin{bmatrix}
        \mm {J}_{\mm H, \mm H^*}& \mm {J}_{\mm H^*, \mm H^*}\\
        \mm {J}_{\mm H, \mm H}& \mm {J}_{\mm H^*, \mm H}
    \end{bmatrix},
    \\
         &\mm {J}_{\mm H, \mm H}= \mm {J}_{\mm H^*, \mm H^*}^* = L\mm K \left[ (\mm H^H \mm R_{y, j}^{-1})^T \otimes (\mm H^H \mm R_{y, j}^{-1}) \right],
        \\
         &\mm {J}_{\mm H, \mm H^*}= \mm {J}_{\mm H^*, \mm H}^* = L (\mm H^H \mm R_{y, j}^{-1}\mm H)^T \otimes \mm R_{y, j}^{-1}.
    \end{aligned}
\end{equation}

Then, we apply the chain rule, which gives the final FIM:
\begin{align}
\label{eq:calProc}
    & \mathbf{J}(\bm \xi) = \mm P\mm \Gamma \mm J(\mm H, \mm H^*) \mm \Gamma^H \mm P^T, \\
    & \mm \Gamma =  \begin{bmatrix}
        \mm I_{2MC} & \mm I_{2MC}\\
        -j \mm I_{2MC}& j\mm I_{2MC}
    \end{bmatrix}, \\
    & \mm P = \begin{bmatrix}
        \mm I_{C} \otimes \mm C_1 & \mm 0 & \mm I_{C} \otimes \mm C_2 & \mm 0\\
        \mm 0 & \mm I_{C} \otimes \mm C_1 & \mm 0& \mm I_{C} \otimes \mm C_2
    \end{bmatrix}.
\end{align}

Substitute the four sub-blocks in (\ref{eq:calProc}) into (\ref{eq:kappadef}), we can obtain the results in \textit{proposition 2}:
\begin{align}
    \kappa &= 
    \bm \xi_t^T\left[\left.\mathbf{J}_{\bm \xi_t \bm \xi_t}\right |_{\bm \xi_t = \mm 0}
    -\left.\mathbf{J}_{\bm \xi_t \bm \xi_d} \right |_{\bm \xi_t = \mm 0}
    \mathbf{J}_{\bm \xi_d \bm \xi_d}^{-1} |_{\bm \xi_t = \mm 0}
    \left.\mathbf{J}_{\bm \xi_d \bm \xi_t }\right |_{\bm \xi_t = \mm 0}\right]\bm \xi_t \notag \\
    &= 
    \begin{bmatrix} 
    \mstr{vec}^T (\tilde{\mv H}_t) & \mstr{vec}^T (\tilde{\mv H}_t^*)
    \end{bmatrix} 
    \begin{bmatrix} 
    \mm {J}_{\mm H, \mm H^*} & \mm 0\\ 
    \mm 0 & \mm {J}_{\mm H^*, \mm H}
    \end{bmatrix} 
    \begin{bmatrix} 
    \mstr{vec} (\tilde{\mv H}_t)\\ 
    \mstr{vec} (\tilde{\mv H}_t^*)
    \end{bmatrix} \notag \\
    &= 2L\Re\left[ \mstr{vec}^H (\tilde{\mv H}_t) \mm {J}_{\mm H, \mm H^*}|_{\bm \tilde{\mm H}_t = \mm 0} \mstr{vec} (\tilde{\mv H}_t) \right] \notag \\
    &= \frac{2L}{\sigma_r^2}\mtr\left [ \tilde{\mm H}_t \tilde{\mm H}_d^H (  \sigma_r^2\mm I_M + \tilde{\mm H}_d\tilde{\mm H}_d^H)^{-1} \tilde{\mm H}_d \tilde{\mm H}_t^H \right ].
\end{align}

\subsection{Detector and Performance of Active Detection} \label{active}
In this appendix, we present a conventional active detector and analyze its performance, which serves as the performance upper bound for passive detection considered in this work. 

Consider $M$ SRs equipped with ULAs with $N_r$ elements. The composite hypothesis testing problem is formulated as:
\begin{equation}
\label{eq:HT_act}
    \begin{aligned} 
        \mathcal{H}_0: \quad
        \mm Y_a & = \mm Z_a,\\
        \mathcal{H}_1: \quad
        \mm Y_a & = \mm H_t \mm S + \mm Z_a,\\
    \end{aligned}
\end{equation}
where $\mm S$ is the waveform known to the SRs, $\mm H_t$ is the target channel unknown to the SRs. The log-statistic for this problem can be derived as follows:
\begin{align}  
    \Lambda_a(\mm Y) &= \ln \frac{\max_{\mv H_t} p(\mm Y | \mathcal{H}_1)}{p(\mm Y | \mathcal{H}_0) }\overset{\mathcal{H}_1} {\underset{\mathcal{H}_0} {\gtrless}} \rho, \notag\\
    & = \frac{1}{\sigma_r^2}\mtr(\mm Y\mm Y^H) -  \frac{1}{\sigma_r^2}\max_{\mv H_t} \mtr\left[ (\mm Y - \mv H_t\mm S)(\mm Y - \mv H_t\mm S)^H \right ] \notag\\
    &= \frac{1}{\sigma_r^2}\max_{\mv H_t} \mtr\left[ \mm Y \mm S^H \mv H_t^H + \mv H_t\mm S \mm Y^H - \mv H_t\mm S \mm S^H \mv H_t^H \right ] \notag\\
    \label{eq:act_detector}
    &= \frac{1}{\sigma_r^2}\mtr\left[ \mm Y \mm S^H (\mm S \mm S^H)^{-1} \mm S \mm Y^H\right ],
\end{align}
where $L> C$ is assumed, which is generally satisfied in practical situations. Under $\mathcal{H}_0$, $\Lambda_a(\mm Y)$ follows a central Chi-squared distribution with degree-of-freedom $2MC$. Under $\mathcal{H}_1$, $\Lambda_a(\mm Y)$ follows a non-central Chi-squared distribution \cite{de2007design}, with degree-of-freedom $2MC$ and non-centrality parameter $\kappa_{\text{act}} = \frac{2}{\sigma_r^2}\mtr(\mv H_t\mm S \mm S^H \mv H_t^H) \approx \frac{2L}{\sigma_r^2}\mtr(\mv H_t\mv H_t^H)$, if $L$ is large.

\subsection{Proof of Proposition 3} \label{approx}
We first expand Equation (\ref{eq:kappa}) using the Woodbury formula:
\begin{align}
    \kappa &= \frac{2L}{\sigma_r^4}\mtr\left [ - \frac{1}{\sigma_r^2}\tilde{\mm H}_t^H\tilde{\mm H}_t \tilde{\mm H}_d^H \tilde{\mm H}_d ( \mm I_M + \tilde{\mm H}_d^H\tilde{\mm H}_d/\sigma_r^2)^{-1}\tilde{\mm H}_d^H \tilde{\mm H}_d \right. \notag\\
    & \qquad \left. + \tilde{\mm H}_t^H\tilde{\mm H}_t \tilde{\mm H}_d^H \tilde{\mm H}_d \right ].
    \label{eq:inverse}
\end{align}

Denote the spectral decomposition of the matrix $\tilde{\mm H}_d^H \tilde{\mm H}_d/\sigma_r^2$ as $\mm V \Sigma \mm V^H$, where $\Sigma = \mstr{diag}(\bar{\sigma}_1, \cdots, \bar{\sigma}_C)$ and the eigenvalues are placed in decreasing order. Then, we have:
\begin{align}
    & \frac{1}{\sigma_r^4}\tilde{\mm H}_d^H \tilde{\mm H}_d ( \mm I_M + \tilde{\mm H}_d^H\tilde{\mm H}_d/\sigma_r^2)^{-1}\tilde{\mm H}_d^H \tilde{\mm H}_d\notag \\
    =& \mm V \Sigma \mm V^H (\mm I_M + \mm V \Sigma \mm V^H)^{-1} \mm V \Sigma \mm V^H \notag \\
    =& \mm V \Sigma(\mm I_M + \Sigma )^{-1} \Sigma \mm V^H \label{eq:Vmatrix}.
\end{align}

With (\ref{eq:inverse}) and (\ref{eq:Vmatrix}), we can obtain:
\begin{align}
    \kappa=& \frac{2L}{\sigma_r^2}\mtr\left [ \tilde{\mm H}_t^H\tilde{\mm H}_t \mm V (\Sigma - \Sigma(\mm I_M + \Sigma )^{-1} \Sigma )\mm V^H\right ]\notag \\
    =& \frac{2L}{\sigma_r^2} \sum_{i = 1}^C \frac{\bar{\sigma}_i}{1+\bar{\sigma}_i}\delta_i, \label{eq:refor}
\end{align}
where $\delta_i = \mm v_i^H \tilde{\mm H}_t^H \tilde{\mm H}_t\mm v_i$, $\mm v_i = \mm V[:, i]$ is the $i$-th eigenvector. If $\bar{\sigma}_C \gg 1$, the term $\frac{\bar{\sigma}_i}{1+\bar{\sigma}_i}$ approaches to one, which gives:
\begin{align}
    \kappa \approx \frac{2L}{\sigma_r^2} \sum_{i = 1}^C \delta_i = 2LM \text{SNR}_t.
\end{align}
\vspace{-0.5cm}

\subsection{Relations between optimization problems \texorpdfstring{$(\mathcal P1)$}{P1} and \texorpdfstring{$(\mathcal P2)$}{P2}} \label{equi}
We can relax the objective (\ref{eq:refor_kappa}) through the following steps:
\begin{align}
    \kappa \leq& \frac{2L}{\sigma_r^2} \sum_{i = 1}^C \frac{\bar{\sigma}_1}{1+\bar{\sigma}_1}\delta_i = 2LM\text{SNR}_t \frac{\bar{\sigma}_1}{{1+\bar{\sigma}_1}} \label{eq:ineq_1} \\
    \leq& 2LM\text{SNR}_t \frac{M\text{SNR}_d}{{1+M\text{SNR}_t}} , \label{eq:ineq_2} 
\end{align}
where the fact that $\frac{\bar{\sigma}_1}{1+\bar{\sigma}_1}\geq\frac{\bar{\sigma}_i}{1+\bar{\sigma}_i}$ is used in the first inequality to obtain (\ref{eq:ineq_1}). The following relations are used in the second inequality to obtain (\ref{eq:ineq_2}):
\begin{align}
    M\text{SNR}_d &= \frac{1}{\sigma_r^2}\mtr(\tilde{\mm H}_d\tilde{\mm H}_d^H) = \sum_{i = 1}^C \bar{\sigma}_i \geq\ \bar{\sigma}_1.
\end{align}

Inequality (\ref{eq:ineq_2}) indicates that the non-centrality parameter in the multi-CU case can be relaxed to the same form as that in the single-CU case, as shown in (\ref{eq:kappa2}). If we use this relaxation as the objective, the problem can be written as:
\begin{subequations}
\begin{align}
\label{eq:equi_prob}
    (\mathcal P3): & \quad \max_{\mm W} \quad 2LM\text{SNR}_t \frac{M\text{SNR}_d}{{1+M\text{SNR}_d}}  \tag{70}\\
    \text{s.t.} & \quad (\ref{eq:op_originalsub1}), (\ref{eq:op_originalsub2}). \notag
\end{align}
\end{subequations}

Introduce a variable $t$ and we have the equivalent form:
\begin{subequations}
\begin{align}
    (\mathcal P3.1): & \quad \max_{\mm W, t} \quad 2LM\text{SNR}_t \frac{t}{{1+t}}  \tag{71}\\
    \text{s.t.} & \quad M\text{SNR}_d \geq t, \tag{71a} \\
     & \quad (\ref{eq:op_originalsub1}), (\ref{eq:op_originalsub2}). \notag
\end{align}
\end{subequations}

Considering that the value of $\text{SNR}_d$ is bounded, we can search $t$ to find the optimal solution. Under each fixed $t$, the problem $(\mathcal P3.1)$ is equivalent to $(\mathcal P2)$. Thus, if an optimal threshold value is selected in $(\mathcal P2)$, the solution is equivalent to the relaxed $(\mathcal P1)$.

\vspace{-0.3cm}

}

\bibliographystyle{IEEEtran}
\bibliography{reference}
 
%

 





\end{document}